\documentclass[a4paper,fleqn,usenatbib]{mnras}
\usepackage{newtxtext,newtxmath}
\usepackage[T1]{fontenc}
\usepackage{ae,aecompl}
\usepackage{gensymb}
\usepackage{pdflscape}
\usepackage{graphicx}	
\usepackage{ulem} 
\title[The bumpy, superluminous SN~2017gci]{SN~2017gci: a nearby Type I Superluminous Supernova with a bumpy tail}

\author[A. Fiore et al.]{A. Fiore,$^{1,2}$\thanks{E-mail: achille.fiore@inaf.it} T.-W. Chen,$^{3,4}$ A. Jerkstrand,$^{3}$ S. Benetti,$^{1}$ R. Ciolfi,$^{1,5}$ C. Inserra,$^{6}$
\newauthor
E. Cappellaro,$^{1}$ A. Pastorello,$^{1}$ G. Leloudas,$^{7}$ S. Schulze,$^{8}$ M. Berton,$^{9,10}$
\newauthor
J. Burke,$^{12,13}$ C. McCully,$^{13}$ W. Fong,$^{14}$ L. Galbany,$^{15}$
\newauthor
M. Gromadzki,$^{16}$ C. P. Guti\'errez,$^{11}$ D. Hiramatsu,$^{12,13}$ G. Hosseinzadeh,$^{17}$ 
\newauthor
 D. A. Howell,$^{12,13}$ E. Kankare,$^{18}$ R. Lunnan,$^{3}$ T. E. M\"uller-Bravo,$^{11}$ D. O' Neill,$^{19}$  
 \newauthor
M. Nicholl,$^{20,21}$ A. Rau,$^{4}$J. Sollerman,$^{3}$ G. Terreran,$^{14}$ S. Valenti,$^{22}$ D. R. Young$^{19}$
\\
$^{1}$INAF - Osservatorio Astronomico di Padova, Vicolo dell'Osservatorio 5, I-35122 Padova, Italy\\
$^{2}$Dipartimento di Fisica e Astronomia `G. Galilei', Universit\`a di Padova, Vicolo dell'Osservatorio 3, I-35122 Padova, Italy \\
$^{3}$ Department of Astronomy, Oskar Klein Centre, Stockholm University, Albanova, 10691 Stockholm, Sweden \\
$^{4}$Max-Planck-Institut f{\"u}r Extraterrestrische Physik, Giessenbachstra\ss e 1, 85748, Garching, Germany\\
$^{5}$INFN, Sezione di Padova, Via Francesco Marzolo 8, I-35131 Padova, Italy \\
$^{6}$School of Physics \& Astronomy, Cardiff University, Cardiff, UK \\
$^{7}$DTU Space, National Space Institute, Technical University of Denmark, Elektrovej 327, 2800 Kgs. Lyngby, Denmark\\
$^{8}$ Department of Particle Physics and Astrophysics, Weizmann Institute of Science, Rehovot 7610001, Israel\\
$^{9}$Finnish Centre for Astronomy with ESO (FINCA), University of Turku, Vesilinnantie 5, FI-20014 University of Turku, Finland \\
$^{10}$Aalto University Mets{\"a}hovi Radio Observatory, Mets{\"a}hovintie 114, FI-02540 Kylm{\"a}l{\"a}, Finland \\ 
$^{11}$Department of Physics and Astronomy, University of Southampton, Southampton, Hampshire, SO17 1BJ, UK \\
$^{12}$Department of Physics, University of California, Santa Barbara, CA 93106-9530, USA \\
$^{13}$Las Cumbres Observatory, 6740 Cortona Dr, Suite 102, Goleta, CA 93117-5575, USA \\
$^{14}$Center for Interdisciplinary Exploration and Research in Astrophysics (CIERA) and Department of Physics and Astronomy,\\ Northwestern University, Evanston, IL 60208\\
$^{15}$Departamento de F\'isica Te\'orica y del Cosmos, Universidad de Granada, E-18071 Granada, Spain\\
$^{16}$Astronomical Observatory, University of Warsaw, Al. Ujazdowskie 4, 00-478 Warszawa, Poland\\
$^{17}$Center for Astrophysics | Harvard \& Smithsonian, 60 Garden Street, Cambridge, MA 02138-1516, USA \\
$^{18}$Department of Physics and Astronomy, University of Turku, Vesilinnantie 5, FI-20014 Turku, Finland\\
$^{19}$Astrophysics Research Centre, School of Mathematics and Physics, Queens University Belfast, Belfast BT7 1NN, UK\\
$^{20}$Birmingham Institute for Gravitational Wave Astronomy and School of Physics and Astronomy, University of Birmingham, Birmingham B15 2TT, UK \\
$^{21}$Institute for Astronomy, University of Edinburgh, Royal Observatory, Blackford Hill, EH9 3HJ, UK \\
$^{22}$Department of Physics and Astronomy, University of California, 1 Shields Avenue, Davis, CA 95616-5270, USA \\
}
\date{Accepted XXX. Received YYY; in original form ZZZ}
\pubyear{2015}
\begin{document}
\label{firstpage}
\pagerange{\pageref{firstpage}--\pageref{lastpage}}
\maketitle
\begin{abstract}
We present and discuss the optical spectro-photometric observations of the nearby ($z=0.087$) Type I superluminous supernova (SLSN I) SN 2017gci, whose peak K-corrected absolute magnitude reaches $M_g=-21.5$ mag. Its photometric and spectroscopic evolution includes features of both slow- and of fast-evolving SLSN I, thus favoring a continuum distribution between the two SLSN-I subclasses. In particular, similarly to other SLSNe I, the multi-band light curves of SN 2017gci show two re-brightenings at about 103 and 142 days after the maximum light. Interestingly, this broadly agrees with a broad emission feature emerging around 6520 \AA{} after $\sim$51 days from the maximum light, which is followed by a sharp knee in the light curve. If we interpret this feature as H$\alpha$, this could support the fact that the bumps are the signature of late interactions of the ejecta with a (hydrogen-rich) circumstellar material. Then we fitted magnetar- and CSM-interaction- powered synthetic light curves onto the bolometric one of SN 2017gci. In the magnetar case, the fit suggests a polar magnetic field $B_{\rm p}\simeq6\times10^{14}$ G, an initial period of the magnetar $P_{\rm initial}\simeq2.8$ ms, an ejecta mass $M_{\rm ejecta}\simeq9\,\mathrm{M_\odot}$ and an ejecta opacity $\kappa\simeq0.08\,\mathrm{cm^{2}\,g^{-1}}$. A CSM-interaction scenario would imply a CSM mass $\simeq5\,\mathrm{M_\odot}$ and an ejecta mass $\simeq12\,\mathrm{M_\odot}$. Finally, the nebular spectrum of phase +187 days was modeled, deriving a mass of $\sim10\,M_\odot$ for the ejecta. Our models suggest that either a magnetar or CSM interaction might be the power sources for SN 2017gci and that its progenitor was a massive ($40\,M_\odot$) star.
\end{abstract}
\section{Introduction}
\label{sec:intro}
Superluminous supernovae (SLSNe) were initially defined as those supernovae (SNe) whose peak-absolute magnitude is brighter than -21 mag \citep{galyam2012}. They are intrinsically rare objects often discovered in metal-poor dwarf host galaxies \citep{chenetal2013,lunnanetal2014,leloudasetal2015,perleyetal2016,chenetal2017a,schulzeetal2018}. The origin of such peculiar transients represents a major challenge for contemporary astrophysics since it raises some fundamental questions about the ultimate stages of the evolution of massive stars. From an observational point of view, SLSNe can be broadly classified according to their hydrogen abundance. SLSNe~I are H poor, although some of them display a late ($\gtrsim100$ days) occurrence of H$\alpha$ \citep[a fraction estimated to be $\sim15\%$,][]{yanetal2017}, while Type II SLSNe display Balmer lines in their optical spectra. Recently it has been proposed that $M_g=-19.8\,\mathrm{mag}$ can be used as a luminosity threshold for the SLSNe~I subclass only \citep{galyam2018b}. However, this does not seem to correspond to a sharp edge in the luminosity function of H-poor SNe \citep{deciaetal2018,galyam2018b,quimbyetal2018} and the SLSN~I classification is generally inferred with a spectrum taken at about the maximum luminosity. This is characterized by a hot blue continuum (with a blackbody temperature $T_\mathrm{BB}\simeq10000-15000$ K) with O {\scriptsize II} absorptions between $3000-5000$ \AA{}.

Determining which physical mechanisms drive the explosion of a SLSN is not obvious. Therefore the discovery of nearby SLSNe (with $z\lesssim0.1$) is of particular interest since it may allow for higher resolution spectra, possibly in a wider wavelength range. A handful of viable scenarios have been invoked to explain the luminosity of SLSNe, as e.~g. the onset of the pair-instability mechanism \citep[e~g.][]{yoshidaetal2016} in very massive stars \citep[heavier than $\sim130\,\mathrm{M_\odot}$,][]{rakavyandshaviv1967,galyametal2009}. In such a scenario, the central pressure drop caused by the $e^{+},e^{-}$ pair creation promptly triggers the collapse of the star and the thermonuclear explosion of its core with an overwhelming production of nickel. Nonetheless, the amount of $^{56}$Ni mass required for an absolute peak-magnitude brighter than $\sim-21$ mag could make the rise time of the light curves (LCs) too slow \citep{nicholletal2013} compared to the observations. Moreover, the spectra of the slow SLSNe~I cannot be fitted by pair-instability models \citep{dessartetal2013,jerkstrandetal2016}. Another possibility lies in the interaction of the SN ejecta with a circumstellar material \citep[CSM, e. g. ][]{chevalierandfransson2003,chevalierandirwin2011,ginzburgandbalberg2012,chatzopoulosetal2012,chatzopoulosetal2013,nicholletal2014,chenetal2015} which was lost by the progenitor star, e~g., via stellar winds or during a pulsational pair-instability phase. If so, the SN ejecta crashes into surrounding shells or clumps of dense matter and drives a shock at the collision edge. This can convert the kinetic energy of the SN ejecta to radiation. However, there are generally no `standard' spectroscopic signatures \citep[i. e. narrow emission lines, as in the case of Type IIn (SL)SNe, e~g. SN 2006gy,][]{smithetal2007} of CSM interaction in the spectra of SLSNe~I \citep{lunnanetal2019}. On the other hand, the presence of the intermediate-width Mg {\scriptsize II} resonance doublet around $\sim2800$ \AA{} \citep{lunnanetal2018}, the late broad H$\alpha$ emission \citep{yanetal2015} and the LC oscillations (bumps) of some SLSNe I \citep{nicholletal2015,yanetal2017} strongly support that the interaction with CSM must be taken into account.

Finally, a model which has growing consensus within the astrophysical community considers that the luminosity of SLSNe~I is sustained by the spin-down radiation of a nascent magnetar \citep[e.~g.][]{kasenandbildsten2010,woosley2010,suzukiandmaeda2017,suzukiandmaeda2019}. According to this scenario, a highly-magnetized, newly-born neutron star is the compact remnant left by the SLSN explosion. Similarly to the case of a pulsar-wind nebula \citep[e~g.][]{metzgeretal2014}, the energy radiated by the neutron star via magnetic-dipole braking inflates a low-density, radiation-dominated photon-pair plasma nebula that afterwards thermalizes into the expanding ejecta, thus acting as a (possibly dominant) power source to explain the luminosity of SLSNe~I. The magnetar scenario is favoured also by the association of the superluminous SN 2011kl \citep{greineretal2015} with an ultra long gamma ray burst. Initially, it was proposed that SLSNe~I might share the environment with fast radio bursts (FRBs) \citep{nicholletal2017,margalitetal2018} but the recent discovery of two FRBs with a massive host galaxy \citep{ravietal2019,marcoteetal2020} disfavours this association.

SLSNe~I are actually a heterogeneous class of transients. In fact, it is possible to distinguish between at least two subclasses, depending on whether their LCs evolve in a slow or a fast fashion. Slow-evolving SLSNe~I have a rise time towards the maximum luminosity which exceeds 50 days, whereas the fast-evolving SLSNe~I reach the maximum light in less than 30 days. Although a continuum distribution likely fills the gap between the two subclasses \citep{nicholletal2015, deciaetal2018}, the distinction between fast- and slow-evolving SLSNe~I is still used \citep[e. g.][]{kumaretal2020} and helpful to distinguish different rise or decline timescales within the SLSN~I class. In addition, slow-evolving SLSNe~I more often show bumps in their LC both before and after the maximum-luminosity epoch  \citep[][]{inserraetal2017,inserra2019}.

SN~2017gci is located at $\mathrm{RA}=06^\mathrm{h}\,46^\mathrm{m}\,45.02^\mathrm{s}$ and $\mathrm{Dec}=-27\degree\,14'\,55.8''$ (J2000). It was discovered by Gaia on the August 16th, 2017 \citep{delgado2017} as an apparently hostless, blue transient and named Gaia17cbp. Initially, it was classified as a Cataclysmic Variable-candidate. Later it was reclassified as SLSN~I \citep{lymanetal2017} by the extended Public ESO Spectroscopic Survey for Transient Objects \citep[ePESSTO,][]{smartt2015}. The last $g^{\prime},r^{\prime},i^{\prime},z^{\prime} ,J,H,K_\mathrm{s}$-band imaging frames (taken on September 29th, 2019) show that the host-galaxy flux contribution of SN~2017gci is not completely negligible at optical/NIR wavelengths ($g_{\rm host}\simeq22.8$ mag, $r_{\rm host}\simeq22.2$ mag, $i_{\rm host}\simeq22$ mag, $J_{\rm host}\simeq21.6$ mag, $H_{\rm host}\simeq21.5$ mag, see Section \ref{sec:phot}).

We hereby present the LCs and the spectra of the SLSN~I SN~2017gci. The observations will be made public via WiseRep\footnote{\texttt{https://wiserep.weizmann.ac.il/search/} .}. In addition, we provide an interpretation of the data both with a semi-analytic magnetar-powered modelling and by means of the single-zone SUMO models \citep{jerkstrandetal2017} for the nebular spectra of SLSNe~I. Hereafter, in Section 2 we describe and discuss the photometric observations; Section 3 deals with the spectra of SN~2017gci; in Section 4 we compare the spectra and the LCs of SN~2017gci with those ones of other SLSNe~I and we provide our interpretation of this event within the magnetar scenario; finally we summarize our conclusion in Section 5. Throughout the paper we assume a flat Universe with $\Omega_\mathrm{m}=0.31$ and $H_0=71\pm3\,\mathrm{km\,s^{-1}\,Mpc^{-1}}$. Given such cosmological parameters and a redshift $z=0.0873\pm0.0003$ (see Section \ref{sec:spectra}), we found a luminosity distance for SN~2017gci of $d_\mathrm{L}=392.5^{+23.5}_{-15.9}\,\mathrm{Mpc}$, corresponding to a distance modulus $\mu=37.96\pm0.1\,\mathrm{mag}\,$. Moreover we assume no extinction from the host galaxy since no narrow absorption interstellar line of the Na {\scriptsize  I}{D} doublet \citep{poznanskietal2012} is seen in the optical spectra.
\begin{figure*}
\centering
\includegraphics[width=11 cm]{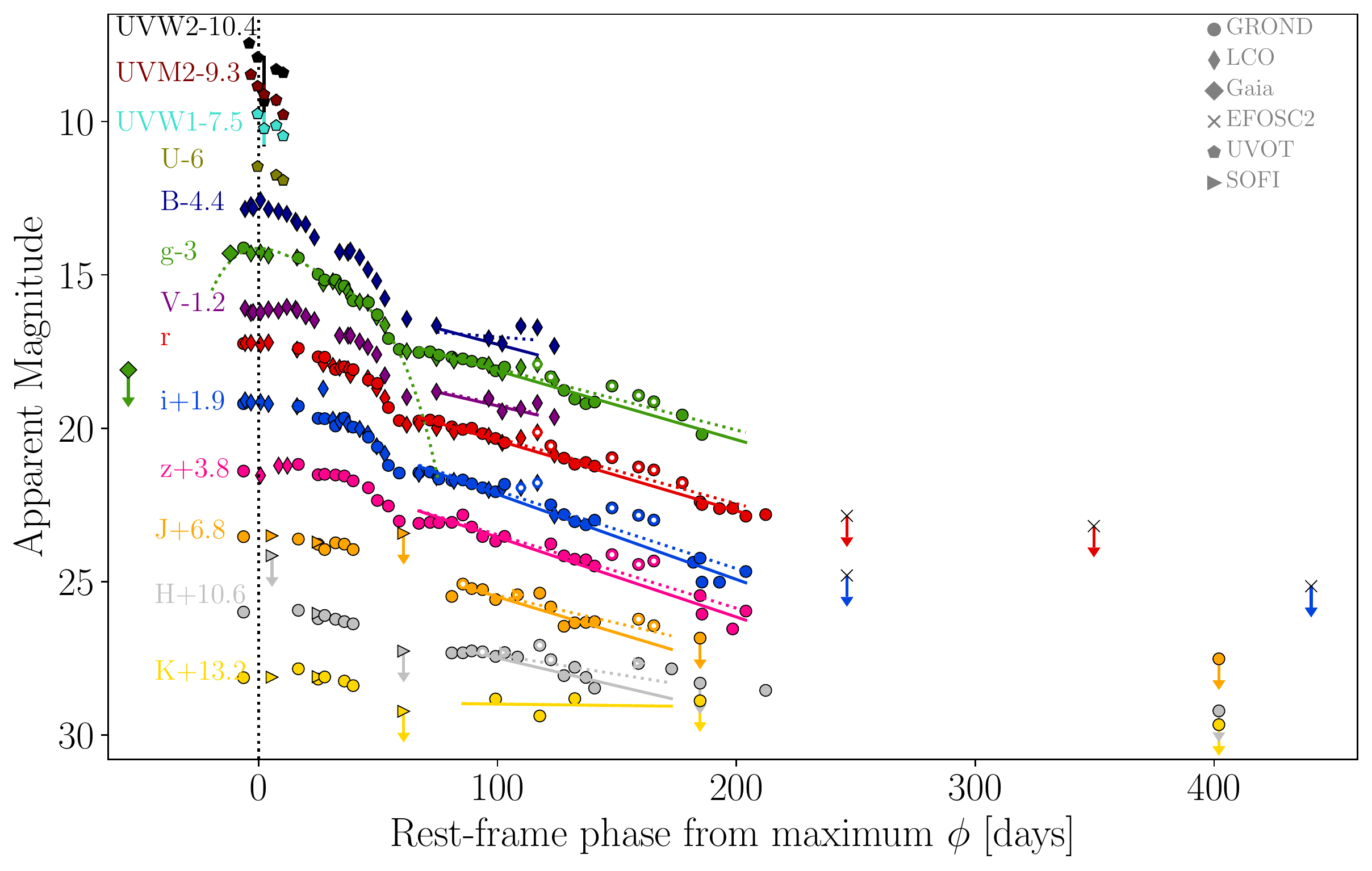}
\caption{S-corrected LCs of SN~2017gci in $UVW2,UVM2,UVW1,U,B,g,V,r,i,z,J,H,K_\mathrm{s}$ bands, respectively plotted in black, brown, cyan, dark green, dark blue, green, purple, red, blue, magenta, orange, silver and yellow. Magnitudes obtained with different instruments were plotted with different symbols, as labelled in grey in the upper-right corner. {The green dotted line represents  a {4th-order} polynomial fit of the early LC to estimate the maximum epoch in $g$ band}. Dotted lines represent the linear fit {to the data with} rest-frame phases later then 74 days, while solid lines represent the linear fits once the bumps have been excluded as explained in the text. The latter points are plotted as empty dots. Arrows correspond to $2.5\sigma$ detection limits. Magnitudes are in AB system. }
\label{fig:lc}
\end{figure*}
\begin{figure*}
\centering
\includegraphics[width=11cm]{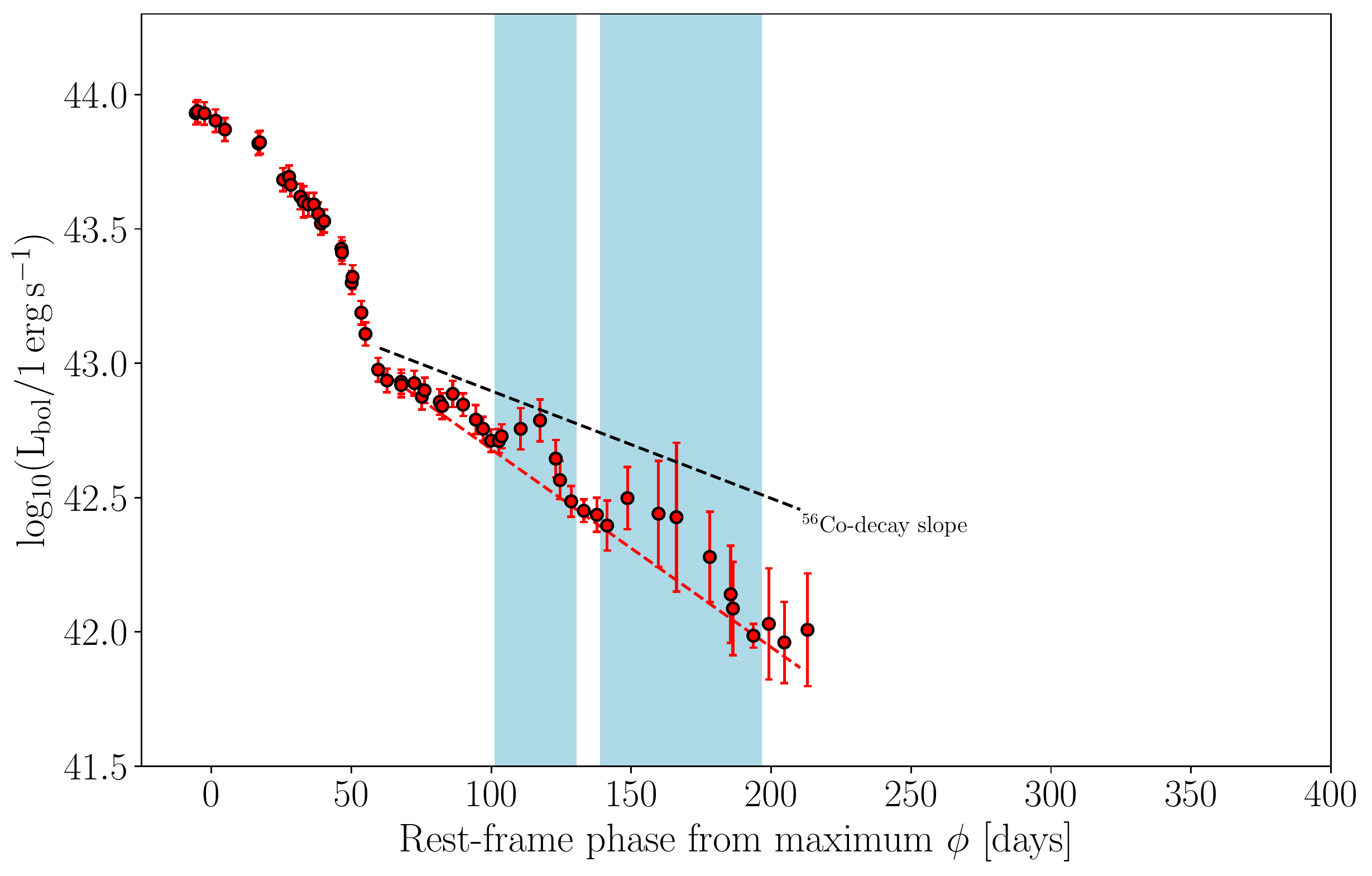}
\caption{Pseudo-bolometric LC of SN~2017gci { (computed after having applied the S-corrections and the K-corrections to multi-band photometry, see text)}. Red dots: pseudo-bolometric LC obtained integrating the SED with the trapezoidal rule. Luminosities are in logarithmic scale and arrows correspond to 2.5$\sigma$ limits. The light-blue shaded areas refer to the epochs during which the bumps occur.}
\label{fig:blc}
\end{figure*}
\section{Photometry}
\label{sec:phot}
\subsection{Observations and preliminary reduction}
\label{sec:phot1}
We performed most of the photometric follow-up with the MPG 2.2m telescope+GROND \citep[Gamma-Ray Burst Optical and Near-Infrared Detector,][]{greineretal2008} as a part of GREAT survey \citep{chenetal2018} and with NTT+EFOSC2 \citep{buzzonietal1984}. Pre maximum- and maximum-epoch data are scarce, but some epochs near the peak were obtained thanks to the photometry of the Las Cumbres Observatory\footnote{\texttt{https://lco.global/} .} (LCO) Global Telescope network. These observations were obtained with the camera Sinistro \citep{brownetal2011} built for the 1m-class LCO telescopes. The set of photometric data we have collected consists of $g^{\prime},r^{\prime},i^{\prime},z^{\prime},J,H,K_\mathrm{s}$-band images taken at ESO La Silla Observatory with 2.2m+GROND, $B,V,g,r,i,z$-filter images taken at LCO, $UVW2,UVM2,UVW1,U,B,V$-filter images obtained with the {{\textit{Swift}} }Ultraviolet/Optical Telescope (UVOT) and $J,H,K_\mathrm{s}$-filter frames obtained with NTT+SOFI \citep[Son OF Isaac, ][]{moorwoodetal1988}. To pre-reduce the EFOSC2 frames, we applied standard overscan, bias and flatfielding procedures within IRAF. The SOFI frames were pre-reduced with the PESSTO pipeline \citep{smartt2015}. The GROND images were pre-reduced by the GROND pipeline \citep{kruehleretal2008}, which applies de-bias and flat-field corrections, stacks images and provides astrometry calibration. 
\subsection{Data reduction}
We corrected the $i$ and $z$ EFOSC2 frames for the fringing pattern by means of fringing masks. These were created by downloading and reducing $\sim100$ archival $i$ and $z$ images from the ESO Archive Science Facility\footnote{\texttt{http://archive.eso.org/} .} for each filter at random coordinates, and selecting those with an exposure time $\gtrsim100\,\mathrm{s}$. We took the median of all of them in order to get rid of the field stars present in the frames. After subtracting the median value from each averaged image we obtained the master fringing mask to be subtracted to the frames.

{$B,g,V,r,i,z,J,H,K_{\rm s}$-filter magnitudes were measured using the SNOoPY package \citep{cappellaro2014} with the Point Spread Function (PSF)-fitting technique, via the DAOPHOT tool \citep{stetson1987}. Within this method, a reference PSF is obtained by averaging those ones of isolated field stars and then fitted onto the SN to obtain the instrumental magnitude. Meanwhile, the background underneath the SN can be estimated interpolating a low order polynomial to the surrounding regions. In alternative, we removed the host galaxy contribution with the template-subtraction technique, which was also performed within SNOoPY, and via the \texttt{hotpants} package \citep{becker2015}. The template-subtraction method envisages the subtraction of the scientific frames with a template image of the same field taken with the same filter when the SN is absent. After the template subtraction, the magnitudes {are} always derived with the PSF method in the residual frame. We found that the template subtraction method gives indeed more reliable photometric measurements, especially when the SN flux becomes fainter. For SN 2017gci this happens at $\phi\sim$100 days after maximum. In Tables~\ref{tab:uvottab}, \ref{tab:griztab}, \ref{tab:bvtab}, \ref{tab:jhktab} if not differently stated, the reported magnitudes have been derived after template subtraction.}

The $g,r,i,z$-template frames were downloaded from the Image Cutout Server\footnote{ \texttt{https://ps1images.stsci.edu/cgi-bin/ps1cutouts/} .} of the second Data Release of Pan-STARRS as stack images. Deep $B,V$-template frames were requested to LCO which observed the field of SN~2017gci on 2019 October, 4th (corresponding to {708} rest-frame days after maximum). {For the $J,H,K_{\rm s}$ template frames we used the combination of the last GROND $J,H,K_{\rm s}$-band frames taken on 2019 September, 25th ({700 rest-frame days after the maximum}) and 29th ({703 rest-frame days after the maximum}), assuming that at these very late epochs SN~2017gci faded {well} below the detection limit. Since the host galaxy is not visible in the deep frame taken about 2 year after explosion, we estimated an upper limit for the $K_{\rm s}$ magnitude ($K_{\rm host,uplim}\simeq18.6$ mag) of the host galaxy using the PSF technique.} {Hence we decided to use $B,g,V,r,i,z,J,H$ template-subtracted magnitudes and $K$ PSF magnitudes.} 

{$B,g,V,r,i,z$ magnitudes were calibrated on the field stars identified with the Pan-STARRS \citep[Panoramic Survey Telescope and Rapid Response System,][]{chambersetal2016} catalogue. {The calibration was performed after having applied the color correction \citep[see equation {6} in][]{tonryetal2012} between Pan-STARRS and SDSS filters}. For the $B,V$ images the calibration was done after having converted {the Pan-STARRS magnitudes to Sloan as before, and then} the Sloan magnitudes to Johnsons system following \citet{chonisandgaskell2008}. The NIR magnitudes were instead calibrated with a local sequence of stars from the Two Micron All Sky Survey \citep[2MASS,][]{skrutskieetal2006}.}\\
To measure $UVW2,UVM2,UVW1,U,B,V$ {\textit{Swift}/UVOT} magnitudes we stacked the layers of the individual observing segments with the task \texttt{uvotimsum} and measured the brightness using 5\arcsec-radius aperture with the task \texttt{uvotsource} in HEASoft version 6.25 \citep{heasoft2014a}.\\



{Since we have used several instruments to collect the photometry of SN~2017gci, each one defining its own photometric system, it is necessary to convert all of them into a standard one. The procedure involved is sometimes called S-correction \citep{stritzingeretal2002} and we applied it following the method described in \citet{eliasrosaetal2006} and \citet{pignataetal2004}. Therefore we computed synthetic photometry using the observed-frame spectra by means of the library \texttt{pysynphot}\footnote{\texttt{https://pysynphot.readthedocs.io/} .} both for the standard photometric systems ($m_{s,\rm standard}$) and for the instrumental filters ($m_{s,\rm instr}$)\footnote{{Instrumental transmission functions for the different instruments were retrieved from \texttt{http://svo2.cab.inta-csic.es/theory/fps3/}.}}. For each instrument and each bandpass filter, the S-correction $S_{\rm corr}$ was then computed as $S_{\rm corr}=m_{s,\rm standard}-m_{s,\rm instr}$. We linearly interpolated over the spectroscopic epochs the $S_{\rm corr}$ grid to match the photometric epochs and then we applied the corresponding correction. We estimated a mean statistical uncertainty for this correction by looking at the dispersion around the interpolation and we assumed it to be 0.02 mag. This uncertainty was eventually summed in quadrature with the photometric one to have the final error. However, the above procedure can be performed only when the passband filters are entirely covered by the wavelength range of the spectra. If this is not the case ($U,z,J,H,K_{\rm s}$), we computed the S-correction as before but using blackbody spectral energy distribution reported to the observer frame. We considered two temperature ranges: $T=12000-8000$ K up to 40 days and $T=8000-4000$ K at later phases, broadly corresponding to the blackbody temperatures derived from the SED blackbody fit (see Section~\ref{sec:tempevol}). The maximum of the S-correction computed in the adopted range is taken as a proxy of the S-correction error introduced by the non standard system, which we called $\Delta S_{\rm corr}$. The $\Delta S_{\rm corr}$ values were propagated in our analysis. The reduced $UVW2,UVM2,UVW1,U,B,g,V,r,i,z,J,H,K_\mathrm{s}$ magnitudes are reported in Tab.~\ref{tab:uvottab}, \ref{tab:griztab}, \ref{tab:bvtab}, \ref{tab:jhktab}. The S-corrections $S_{\rm corr}$ and the $\Delta S_{\rm corr}$ values are listed in Tab.~\ref{tab:scorrgrond},\ref{tab:scorrlco},\ref{tab:scorrswift},\ref{tab:scorrerr}. The latter were divided for simplicity in two temperature bins ($4000\,\mathrm{K}<T<8000\,\mathrm{K}$ and $8000\,\mathrm{K}<T<12000\,\mathrm{K}$).}

The S-corrected LCs of SN~2017gci are shown in Fig.~\ref{fig:lc}. Magnitudes are in AB system and the phases are corrected for time dilation (in the following, we will refer to the rest-frame phase with respect to maximum luminosity as $\phi$). {From our photometric data, it is not possible to provide a robust estimate for the maximum luminosity and the corresponding epoch due to a lack of early time coverage}.
To obtain an upper limit on the rise time, we added a non-detection from the Gaia-archival data \citep{gaia2016a,gaia2016b,salgadoetal2017}, whose epoch is June, 27th 2017 (MJD=57931), which was converted to $g$ magnitude\footnote{Useful relationships to convert Gaia magnitudes to those of the standard photometric systems are available in Section 5.3.5 of the Documentation Release (v. 1.2) of the Gaia Data Release 1. This is accessible from the following URL:\\ \texttt{https://gea.esac.esa.int/archive/documentation/GDR1/} .}. Then a {4th-order} polynomial {was} fit over the early $g$-filter magnitudes {allowing us to estimate the epoch and magnitude of maximum luminosity:}
{$\mathrm{MJD}_\mathrm{max}=57990.3^{+8}_{-15}$} for $g_\mathrm{max}=17.1\pm0.3$ mag. 
\begin{figure}
    \centering
    \includegraphics[width=0.54\textwidth]{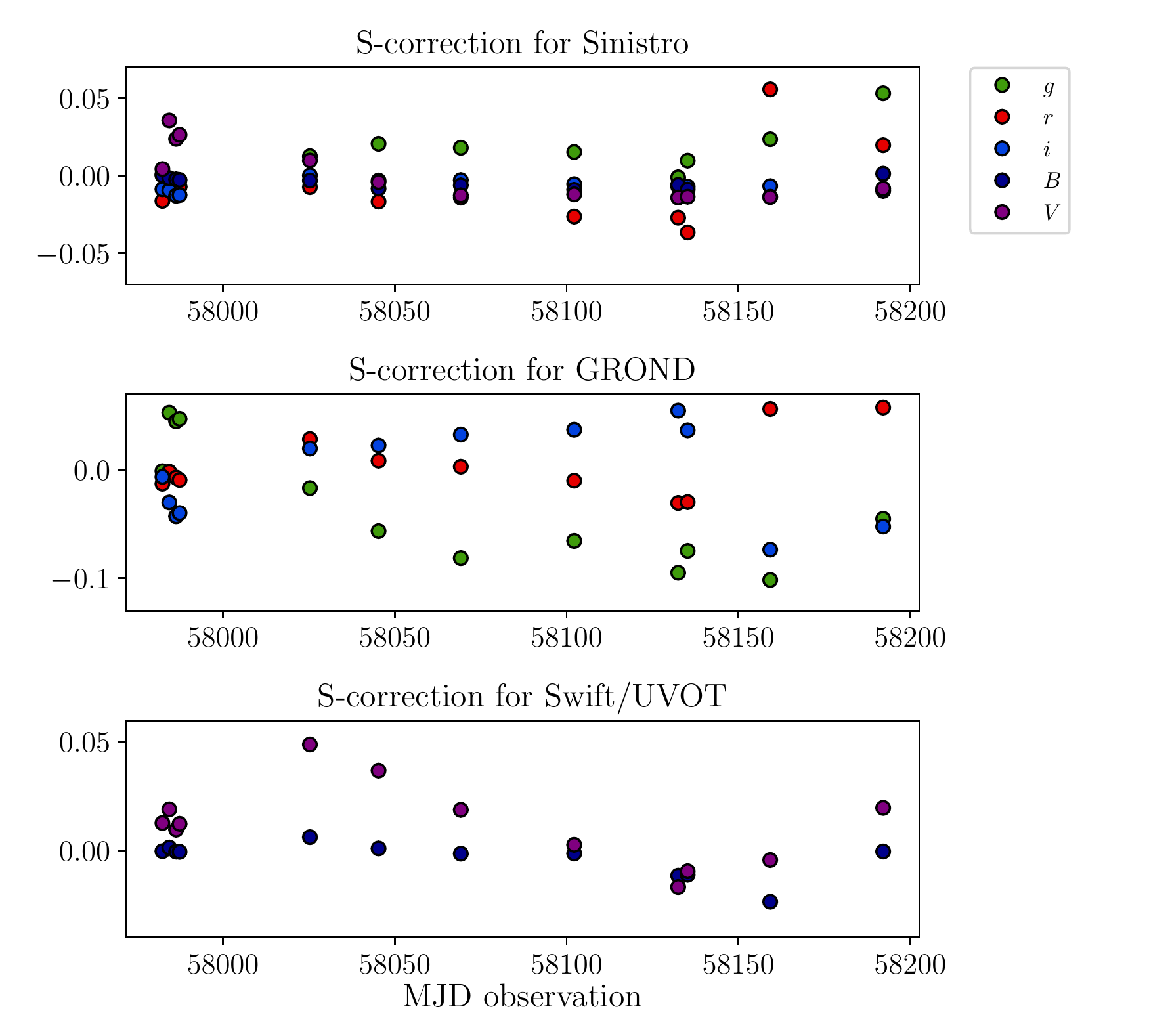}
    \caption{{S-correction for LCO+Sinistro (top panel), GROND+2.2m (middle panel) and \textit{Swift}/UVOT (lower panel). Filled dots are coded as in the label (top right corner).} }
    \label{fig:my_label}
\end{figure}
\subsubsection{K-correction}
\label{sec:kcorr}
For the optical and $J,H$ magnitudes, we obtained the K-corrections from the spectra at our disposal (see Section \ref{sec:spec}). For each of them and for each band-pass filter, we derived a synthetic magnitude via \texttt{pysynphot}. This was done both for the rest-frame spectrum {(for which we computed a synthetic magnitude $m_{s,\rm rest}$)} and for the observed one {(for which we computed a synthetic magnitude $m_{s,\rm obs}$)} via \texttt{pysynphot} distributed via AstroConda\footnote{\texttt{https://astroconda.readthedocs.io/en/latest/} .}. {For each epoch, the K-correction $K$ was computed as $K=m_{s,\rm obs}-m_{s,\rm rest}$. The resulting K-corrections are} listed in Tab.~\ref{tab:kcorr}. Finally, to adjust the sparser time sampling of the spectral epochs to the denser one of the magnitudes we linearly interpolated this table. Similarly, the K-corrections for the $UVW2,UVM2,UVW1,K_\mathrm{s}$-filter magnitudes were estimated by using the SED (retrieved by photometry) in place of the observed spectra.
\subsection{Main characteristics of the LCs and of the bolometric curve}
\label{sec:mclc}
The $g,V,r,i$-filters LCs remain nearly constant for the first $\sim20$ days, while the $B$ {LC} starts to decline earlier (after about {$\sim12$} days from the maximum light). The $U-$ and $UVW2,UVM2,UVW1$-filters LCs {possibly} peak {a} few days before, but the {early-time data coverage is inadequate and so we cannot securely} constrain the maximum-luminosity epoch for those filters. The overall evolution is slower in the $z,J,H,K_\mathrm{s}$ magnitudes, and the early flat phase around the maximum luminosity lasts about $30-40$ days. Then for {$\phi\gtrsim17$} days, the evolution steepens and at {$\simeq54-57$} days the observed LCs present an abrupt change of their slopes. Such a `knee' seems much sharper than the transition region which usually preludes to the so-called `magnetar tail' \citep[see e~g.][]{inserraetal2013,deciaetal2018}. Thereafter, for {$\phi\gtrsim71$} days, the LCs settle on a steady, almost linear decline. During this phase, the LCs display two sharp re-brightenings at {$\phi\sim103$} and {$142$} days. Finally, {after} {$\phi>213$} days, the SN {is no longer detectable}. After this epoch, we took 4 frames in the $g,r,i,z$ bands, 3 in $J$-, 2 in $H$- and 1 in $K_\mathrm{s}$ GROND bands until September 29th, 2019 (see Introduction). However, the template subtracted images provided only $2.5\sigma$ detection limits.
\begin{figure*}
\centering
\includegraphics[width=11 cm]{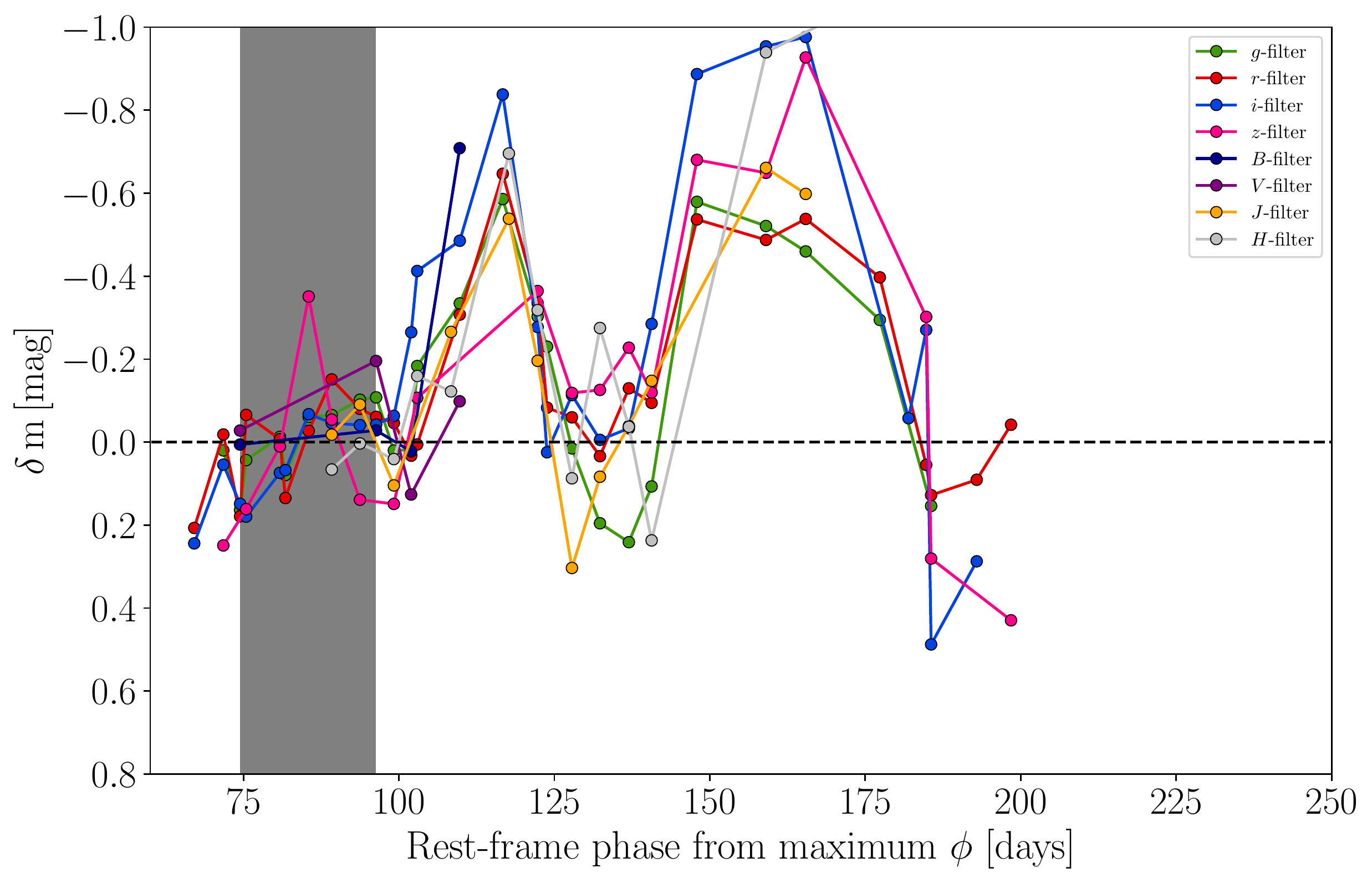}
\caption{Luminosity undulations $\delta m$ in each bandpass filter. Datapoints are coloured as in Fig.~\ref{fig:lc}. The dashed black line marks the $\delta m=0\,\mathrm{mag}$ level. The {gray-}shaded area represents the phase range within which we consider the LC decline not to be affected by the bump features. Therein we computed the standard deviation of the linear-fit residuals to disentangle the bumps (see the text) from the LC decline.}
\label{fig:undul}
\end{figure*} 

Apparent magnitudes were converted to absolute magnitudes once the redshift and the Galactic absorption are known. Given a Galactic extinction $A^{V}_\mathrm{G} = 0.360\,\mathrm{mag}$ \citep{schlafly2011}, the distance modulus $\mu$ and the $g$ apparent magnitude for the maximum given in Section \ref{sec:intro}, the K-corrected absolute peak-magnitude is $M_g=-21.5\pm0.3\,\mathrm{mag}$ in $g$ band. 
We also built the {pseudo-bolometric} LC of SN~2017gci. This was computed by integrating its K-corrected $UVW2,UWM2,UVW1,U,B,g,V,r,i,z,J,H,K_\mathrm{s}$ photometry. We adopted as reference the epochs of the $r$ band photometry, and missing measurements at given epochs for the other filters were obtained through interpolation or, if necessary, by extrapolation assuming a constant colour from the closest available epoch. The fluxes at the filter effective wavelengths, corrected for the Galactic extinction, provide the Spectral Energy Distribution (SED) at each epoch. Then, we integrate the SED with the trapezoidal rule, assuming zero flux at the integration boundaries. 

 We measured the early and late luminosity-decay slopes onto the bolometric LC . The steeper early decline (between {$\phi=30-51$} days from the maximum light) {is estimated} to be {0.040} mag/day, whereas the late one (between {$\phi=60-210$} days from the maximum light) is 0.018 mag/day. {As mentioned above,} a handful of SLSNe~I \citep[e~g. SN 2015bn, ][]{nicholletal2016a} show a bumpy LC. To measure the post-maximum decay slope for {$\phi\gtrsim73$} days, we excluded the bumps from the linear fit. To do that, we proceeded in the following way. For a given wavelength band, we fitted a first-order polynomial using all the magnitudes for {$\phi>73$} days, and for each of those epochs we subtracted the interpolated magnitude to the corresponding observed one. Then, we computed the standard deviation {$\sigma_{73-103}$} of the fit residuals for {$73\lesssim\phi\lesssim\,103\,\mathrm{days}$} (corresponding to the gray shaded area in Fig.~\ref{fig:undul}), since at these epochs the magnitudes do not seem to be affected much by the bumps. At this point we repeated the linear fit, this time excluding all the magnitudes whose difference with the previous fit is brighter than {$1\times\sigma_{73-103}$} mag. Bluer LCs tend to decay faster than the redder ones (except for the $i$ and $z$ bands). Nonetheless, {we specify} that the slope estimates for the {$B,V,J,H,K_\mathrm{s}$} LCs are less accurate because of the evident data paucity. The late-decline slopes were not measured for the $UVW2,UVM2,UWV1,U$-filter LCs since no coeval measure is available. The luminosity excesses in the different bands $\delta m$ over the late post-peak decline rate $\langle m \rangle$ are such that $|\delta m|=|\langle m \rangle-m|\lesssim1\,\mathrm{mag}$ (see Fig.~\ref{fig:undul}).
\begin{figure*} 
\centering
\includegraphics[width=0.7\textwidth]{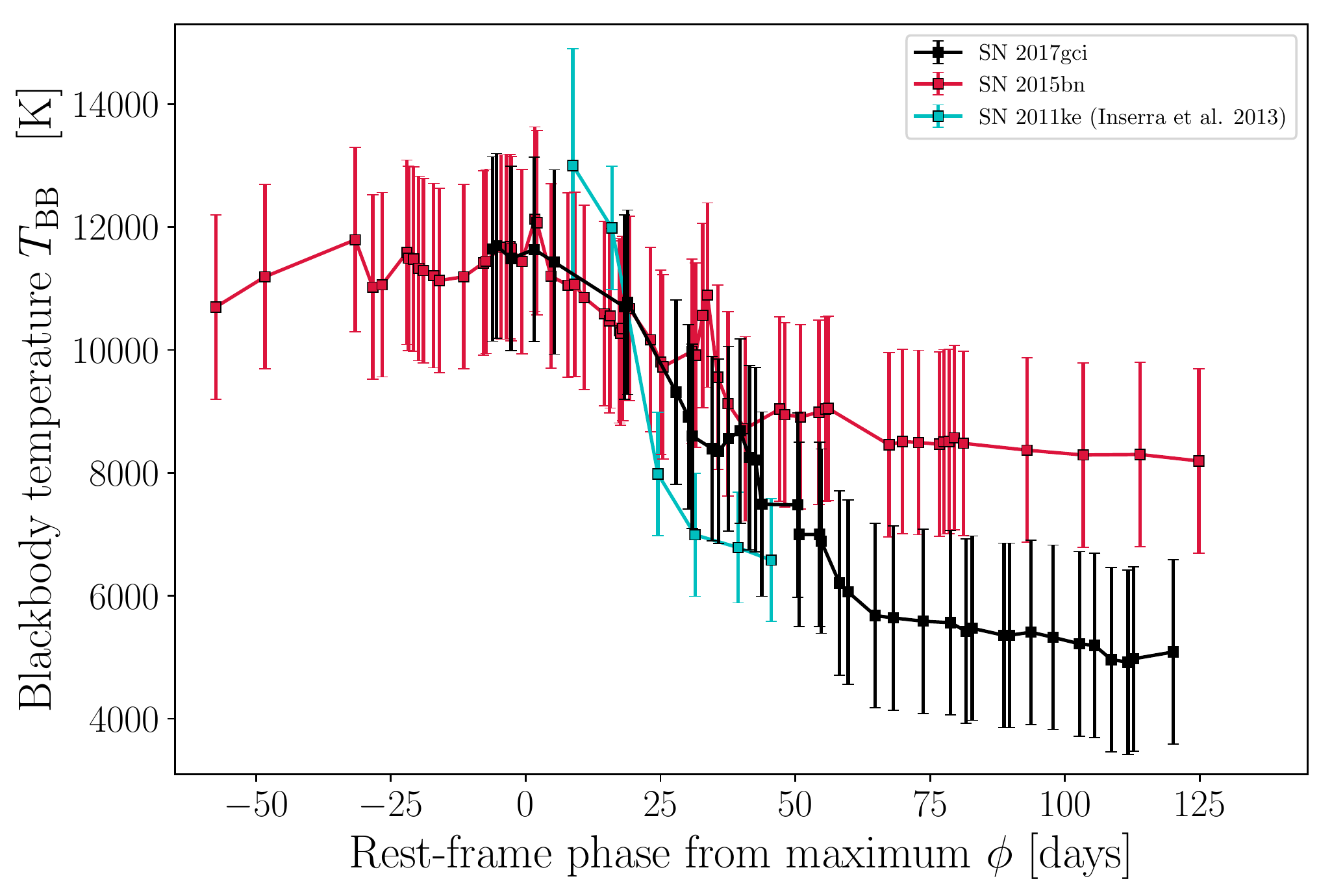}
\caption{Temporal evolution of the black-body temperature {of SN~2017gci (black squares), SN 2015bn (red squares) and SN 2011ke (cyan squares). The blackbody temperatures for both SN~2017gci and SN~2015bn were retrieved by a blackbody fit of the SED, while the temperatures of SN 2011ke are taken from \citet{inserraetal2013}. }   }
\label{fig:bbrad}
\end{figure*}
\section{Spectroscopy}
\label{sec:spec}
\subsection{Observations and data reduction}
Optical spectra were acquired with the ESO New Technology Telescope (NTT)+EFOSC2 at La Silla Observatory, Chile, the ESO Very Large Telescope (VLT)+X-Shooter(XS) {{\citep[see][for a description]{{vernetetal2011}}}} at Paranal Observatory, Chile, the Keck I telescope+LRIS {\citep{okeetal1995}} at W. M. Keck Observatory, Maunakea, Hawaii and the Multiple Mirror Telescope (MMT)+Binospec {\citep{fabricantetal2019}} at Maunakea Observatory. The EFOSC2, Binospec and LRIS spectra were reduced with the standard IRAF tools. {Also, five LCO+FLOYDS spectra were secured (see Tab.~\ref{tab:sfo})}. \\
The two-dimensional raw spectroscopic frames were then corrected for overscan, divided by a normalized flat-field, corrected for cosmic rays \citep[by means of the \texttt{L. A. Cosmic} algorithm, ][]{vandokkum2001}, extracted across the spatial direction after having interpolated the background below the SN with a low-order polynomial fit on the surrounding regions, calibrated in wavelength against HeAr arcs. Then the extracted one-dimensional spectra were calibrated in flux and corrected for telluric absorption thanks to a set of spectrophotometric standard stars. Finally, the flux calibration of these spectra was also checked against the magnitudes retrieved by coeval photometry. The first XS spectrum ({$\phi=187$} days) was reduced following the procedure described in \citet{kruehleretal2014}, whereas the second one ({$\phi=367$} days) was reduced via the \texttt{esoreflex} ESO-pipeline {\citep[v 2.9.1, ][]{freudlingetal2013}}.
\subsection{The spectra}
\label{sec:spectra}
\begin{figure*}
\centering
\includegraphics[width=14 cm]{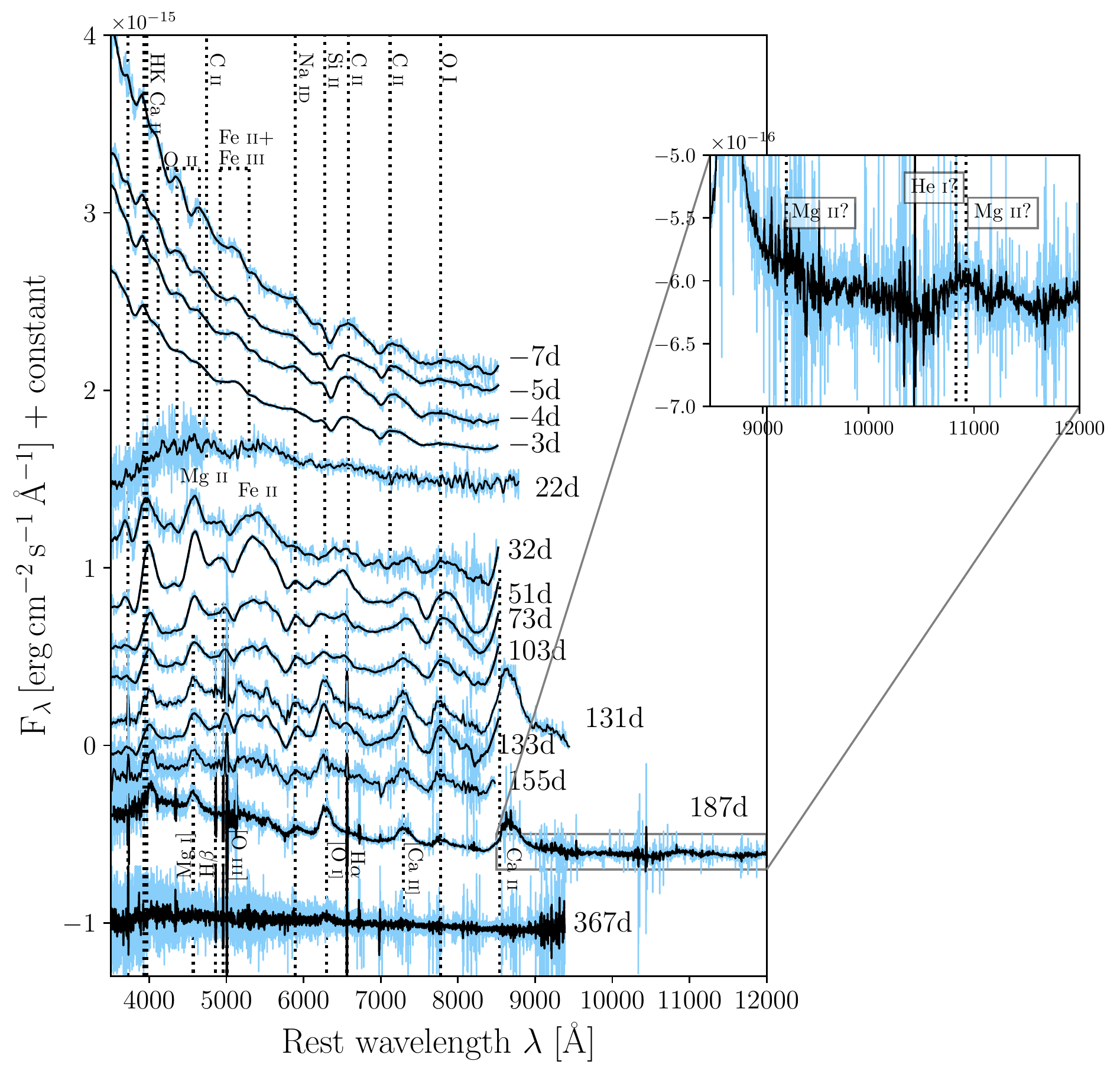}
\caption{Spectral evolution of SN~2017gci. On the right side we plotted the rest-frame phase from the discovery for each spectrum. To provide a clear representation, we smoothed the spectra with a Savitzky-Golay filter. The smoothed spectra (black lines) have been overimposed to the original ones (light-blue lines), and all of them were scaled and offset. The black dashed lines mark the wavelengths at which the spectral features occur in the spectra. For each of them, the ion responsible of the transition is labelled nearby. In the last two spectra (at {$\phi$=187,367} days), the narrow emission lines [O {\scriptsize  III}], H$\alpha$, H$\beta$ were cut up to a certain flux threshold. The epochs of the observations, the scaling factors and the offsets are summarized in Tab.~\ref{tab:sfo}. The insert in the upper-right corner zooms the NIR part of the XS spectrum taken at {187} days.}
\label{fig:spec_evol}
\end{figure*}

The spectral evolution of SN~2017gci is shown in Fig.~\ref{fig:spec_evol}.  In order to increase the signal-to-noise ratio we included in Fig.~\ref{fig:spec_evol} the average between the spectra observed at {$\phi=20$ and $\phi=23$} days which is marked with a phase of $\phi=22$ days in the figure. The identification of the spectral features was done following \citet{howell2017}, \citet{quimbyetal2018}. Until the maximum light, the spectra of SN~2017gci show a hot blue continuum whose black-body temperature reaches $T_\mathrm{BB}\simeq12000-14000\,\mathrm{K}$ in the spectra about the maximum luminosity. On the redder side of the spectra, the broad Na {\scriptsize  I}{D} $\lambda\lambda\,\,5890,5896$ doublet, the C {\scriptsize  II} $\lambda\lambda\,\,6580,7121$ lines and the O {\scriptsize  I} $\lambda\,\,7774$ are evident. Tentatively we also identified the Si {\scriptsize  II} $\lambda\,\,6355$ feature. At shorter wavelengths the doublet H\&K of the Ca {\scriptsize  II} and the W-shaped O {\scriptsize  II} features at $\lambda\lambda\, 4115,\,4357,\,4650$ are also present. To test their identification we compared the pre-maximum spectrum of SN~2017gci at {$\phi=-7\,\mathrm{d}$} with two synthetic spectra computed with TARDIS \citep[Temperature And Radiative Diffusion In Supernovae,][]{kerzendorfandsim2014}, an open-source, Monte Carlo-based, radiative-transfer spectral synthesis code for SN spectra. The first TARDIS spectrum (see Fig.~\ref{fig:3550}, orange dots) was calculated assuming a pure Carbon chemical abundance while in the second one (magenta dots) a pure Oxygen abundance was input. Both of them assume Local Thermodynamic Equilibrium conditions.  As shown in Fig.~\ref{fig:3550}, the features at $\lambda\lambda\,4115,4651$ are well matched by the pure-O spectrum, but we cannot exclude a line blending with C {\scriptsize  II} spectral features. Moreover, we identified two absorptions as the contribution of Fe {\scriptsize  II} + Fe {\scriptsize  III}. The latter is a common feature among the slow SLSNe~I \citep{inserra2019}.
\begin{figure*}
\centering
\includegraphics[width=15 cm]{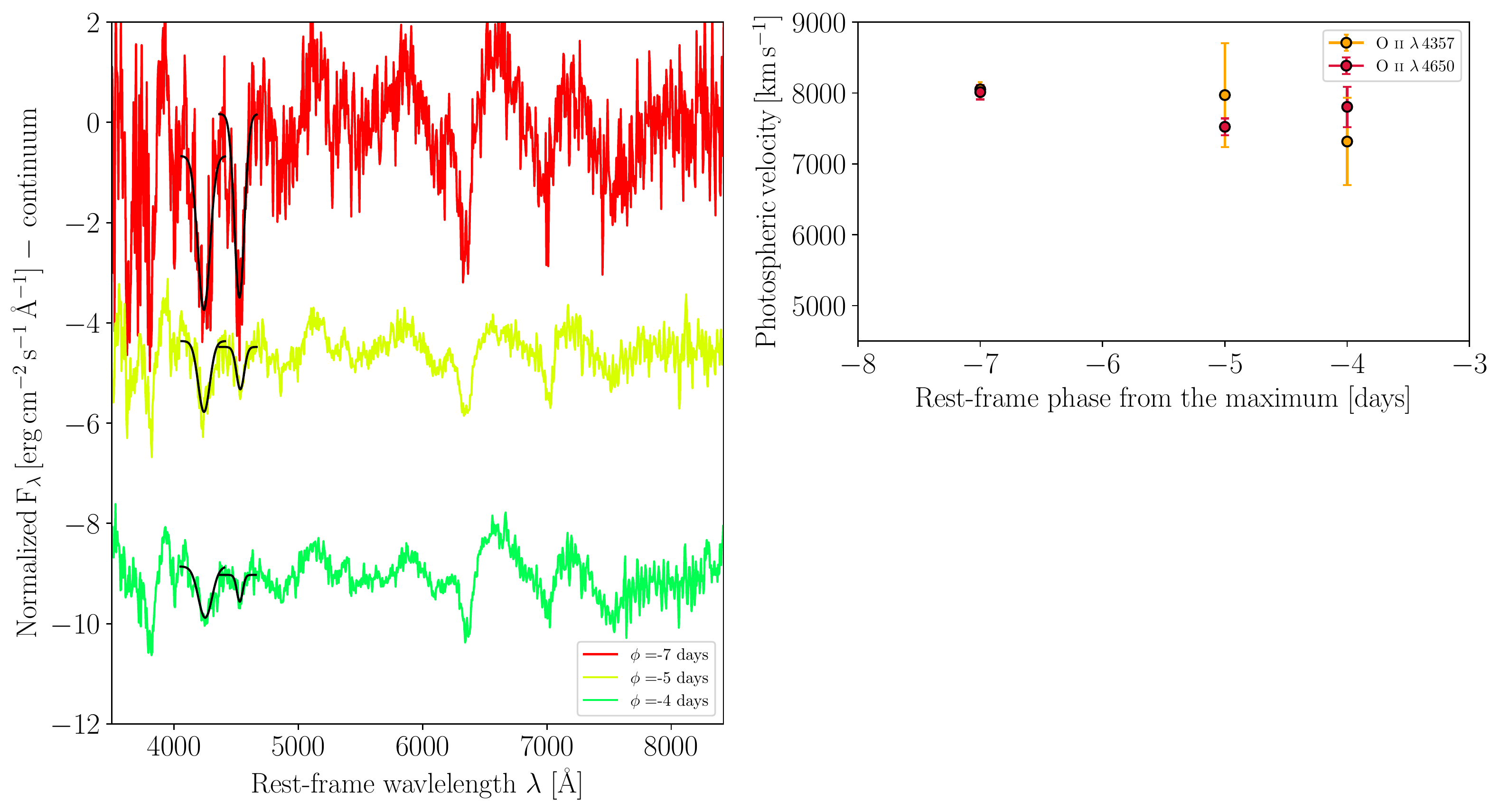}
\caption{{Left panel: normalized and continuum-subtracted spectra between {$\phi=-7$ and $\phi=-4$} days. The O {\scriptsize II} absorption minima were fitted with a gaussian (black solid lines). Right panel: Photospheric velocities retrieved by the absorption minima of the O {\scriptsize II} ($\lambda4357,4650$) features.}}
\label{fig:vphot}
\end{figure*}
{After {$\sim20$} days, the continuum becomes significantly redder with much less prominent O {\scriptsize  II} absorptions. In their place, the Fe {\scriptsize II}  and Mg {\scriptsize  II} features start to be visible.}
From {$\phi\gtrsim33$ days, the spectra of SN~2017gci resemble those of a Type Ic BL SN at maximum luminosity, as expected by a SLSN~I  \citep{pastorelloetal2010}. Then, up to {$\phi\sim156$} there are no significant changes in the the spectra}, except for the continuum becoming even fainter and redder. Surprisingly, at {$\phi\sim51$} days a spectral feature consistent with $\mathrm{H}\alpha$ emerges in the spectra and remains visible until {$\phi\sim133$} days. The occurrence of such a feature {precedes ($\sim3-6$ days)} a LC knee ( see Section~\ref{sec:mclc}).

For ${\phi\geq155}$ days the spectra becomes {`pseudonebular' \citep{nicholletal2019} where emission features start to prevail on the absorption but with a residual fainter continuum}. At these epochs the SN ejecta were cool enough to favour the recombination of the electrons. This reduces the free-electrons density, hence the optical depth and {allow us} to investigate the deepest emitting regions of the SN explosion. Moreover, in such a low-density environment the semi-forbidden and forbidden atomic transitions start to dominate the spectra. In fact, the emissions of the semi-forbidden $\lambda4571$ Mg {\scriptsize  I}], the [O {\scriptsize  I}] doublet at $\lambda\lambda\,6300,6364$ and $\lambda\lambda\,7291,7323$ [Ca {\scriptsize  II}] are present, as well as the strong NIR Ca {\scriptsize  II} $\lambda\lambda\,8498,8542,8662$ triplet. 

Finally, in the {late} spectra {(at $\phi>130$ days)} of SN~2017gci the narrow H$\alpha$- and [O {\scriptsize  III}]- emission lines from the host galaxy become gradually visible. {Using these features} we calculated the redshift of the host galaxy, which turns out to be $z=0.0873\pm0.0003$ (where the uncertainty is derived from the dispersion of the measurements). Moreover, the spectrum at {$\phi=187$} days presents two features between 9000-11000 \AA{} (see the insert in Fig.~\ref{fig:spec_evol}) where the contribution of Mg {\scriptsize II} $\lambda\, 0.92\,\mu\mathrm{m}$ and $\lambda\,1.09\,\mu\mathrm{m}$ He {\scriptsize I} might be involved. He {\scriptsize I} {is not frequently seen among SLSNe I, except in the case of PTF10hgi \citep{quimbyetal2018} and possibly in the case of SN 2012il \citep{inserraetal2013,quimbyetal2018}}. Further interpretation of the {spectrum at {$\phi=187$ days}} will be provided in Section \ref{sec:model}. In the spectrum taken at {$\phi=367$} days almost all the broad features present in the previous spectrum are {no longer present} except a residual contribution from [O {\scriptsize I}]. The NIR part of this spectrum was too faint to be extracted.
\begin{figure}
\centering
\includegraphics[width=8cm]{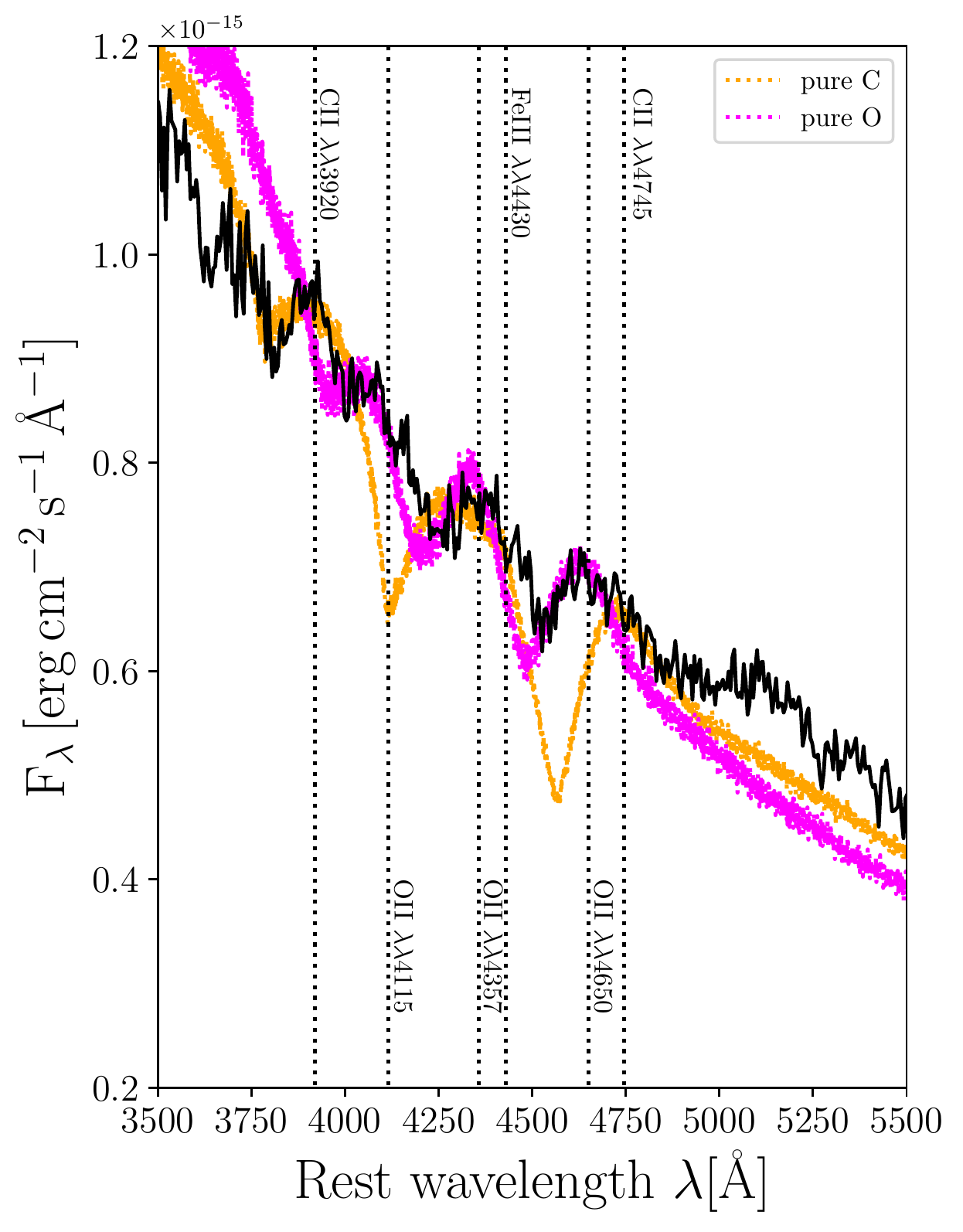}
\caption{The spectral range between $3500-5000$ \AA{}; the black {solid} line is the spectrum of SN~2017gci with rest-frame phase {$\phi=-7$} days. The {dotted} black lines mark the spectral lines which they are labelled with, respectively of Ca {\scriptsize  II}, O {\scriptsize  II}, C {\scriptsize  II} and Fe {\scriptsize  III}. The orange and the magenta dotted lines trace the synthetic profiles, respectively for a pure Carbon and Oxygen composition, computed with TARDIS. For the sake of completeness, we reported also H\&K Ca {\scriptsize  II} doublet, not present in the TARDIS spectrum. The O {\scriptsize  II} features are pretty well matched by the magenta profile.}
\label{fig:3550}
\end{figure}
\begin{figure}
\centering
\includegraphics[width=9cm]{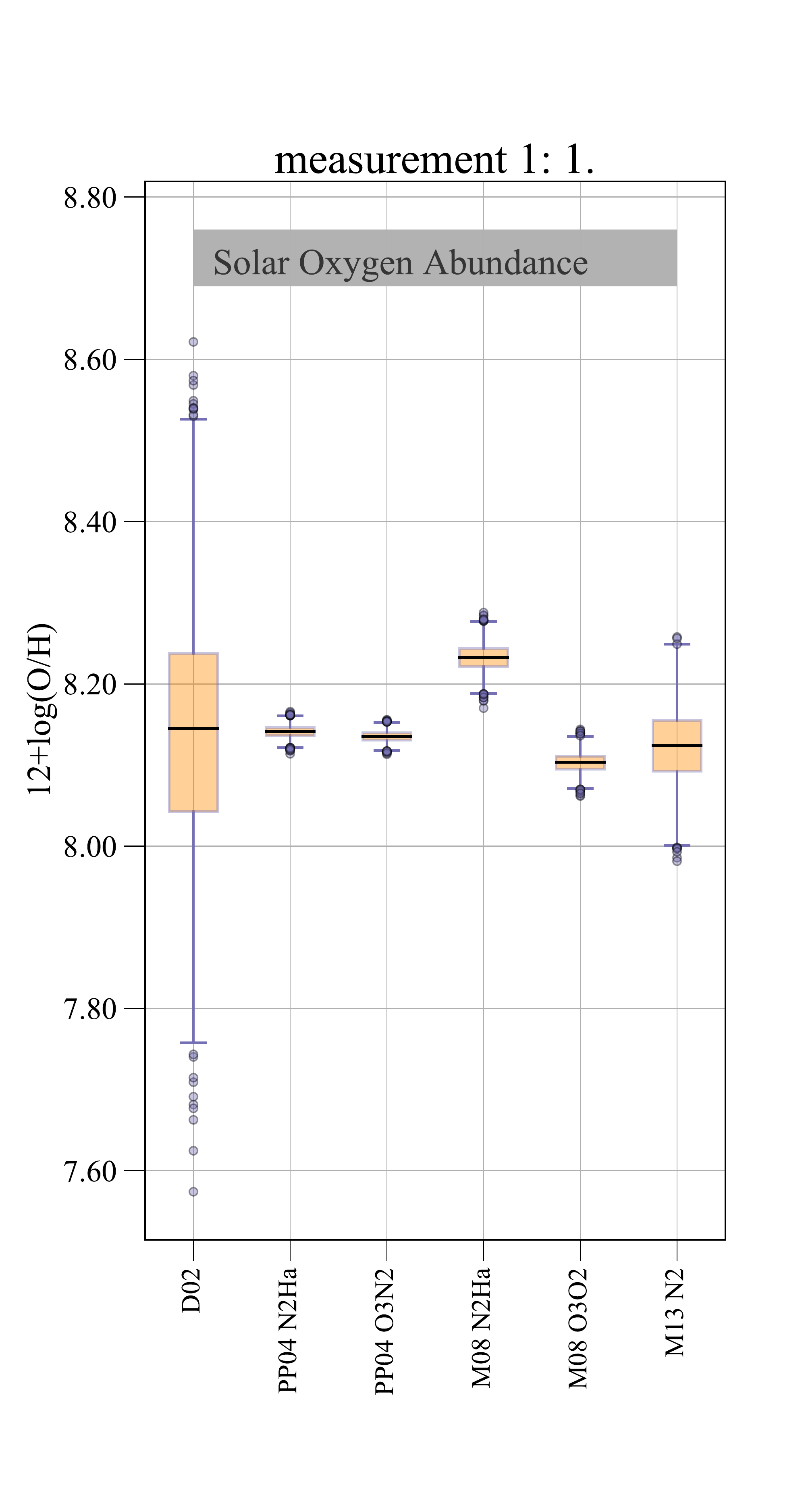}
\caption{A comparison of 6 metallicity estimates at the location of SN~2017gci. Around the nominal metallicity value (black horizontal lines) synthetic data were generated. The orange box reproduce the interquartile ranges (IQRs), and the blue dots are considered outliers since they deviate from the first and third quartile more then $1.5\times\mathrm{IQR}$. The gray box spans over the range of the commonly-used value for the solar oxygen abundance.}
\label{fig:pymcz}
\end{figure}
\begin{figure}
\centering
\includegraphics[width=10 cm]{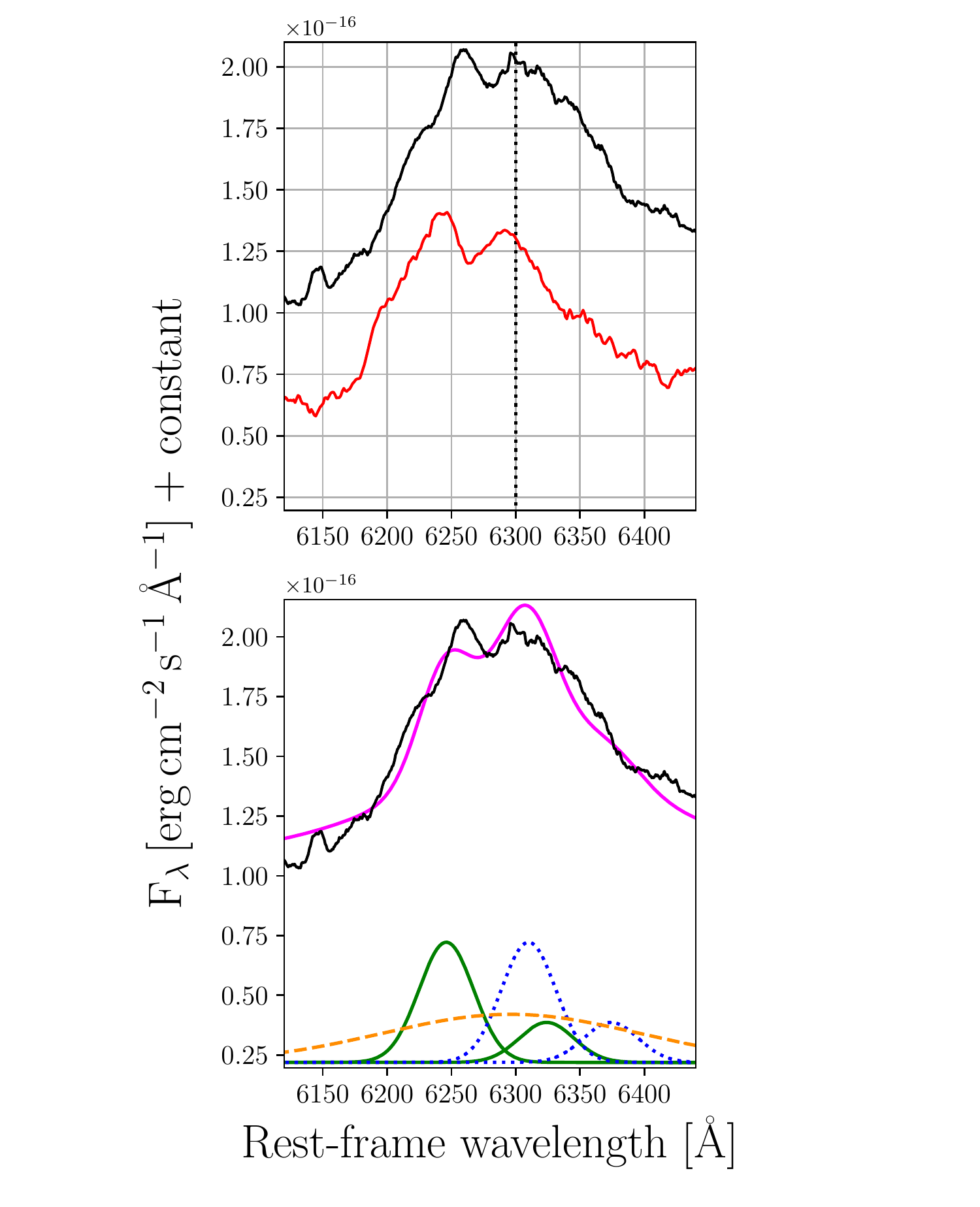}
\caption{Upper panel: comparison of the asymmetric [O {\scriptsize  I}] $\lambda\lambda\,6300,6364$ emission from the spectrum at a rest-frame phase {$\phi=187$} days of SN~2017gci (black solid line) with the same feature shown by the spectrum of the Type Ib SN 2005bf at $\phi=213$ days (red solid line). Lower panel: fit of a composite model (magenta solid line plotted over the observed spectrum) made by five gaussians. A couple of 64 \AA{}-separated gaussians, one at nearly zero velocity (blue dotted line), a couple of gaussian blueshifted of $\sim2000\,\mathrm{km\,s^{-1}}$ with respect to the rest-frame doublet (green solid line) and a broad component which peaks at $\lambda\sim6300$ \AA{} (yellow dashed line).}
\label{fig:oi636364}
\end{figure}
\begin{figure}
\centering
\includegraphics[width=8cm]{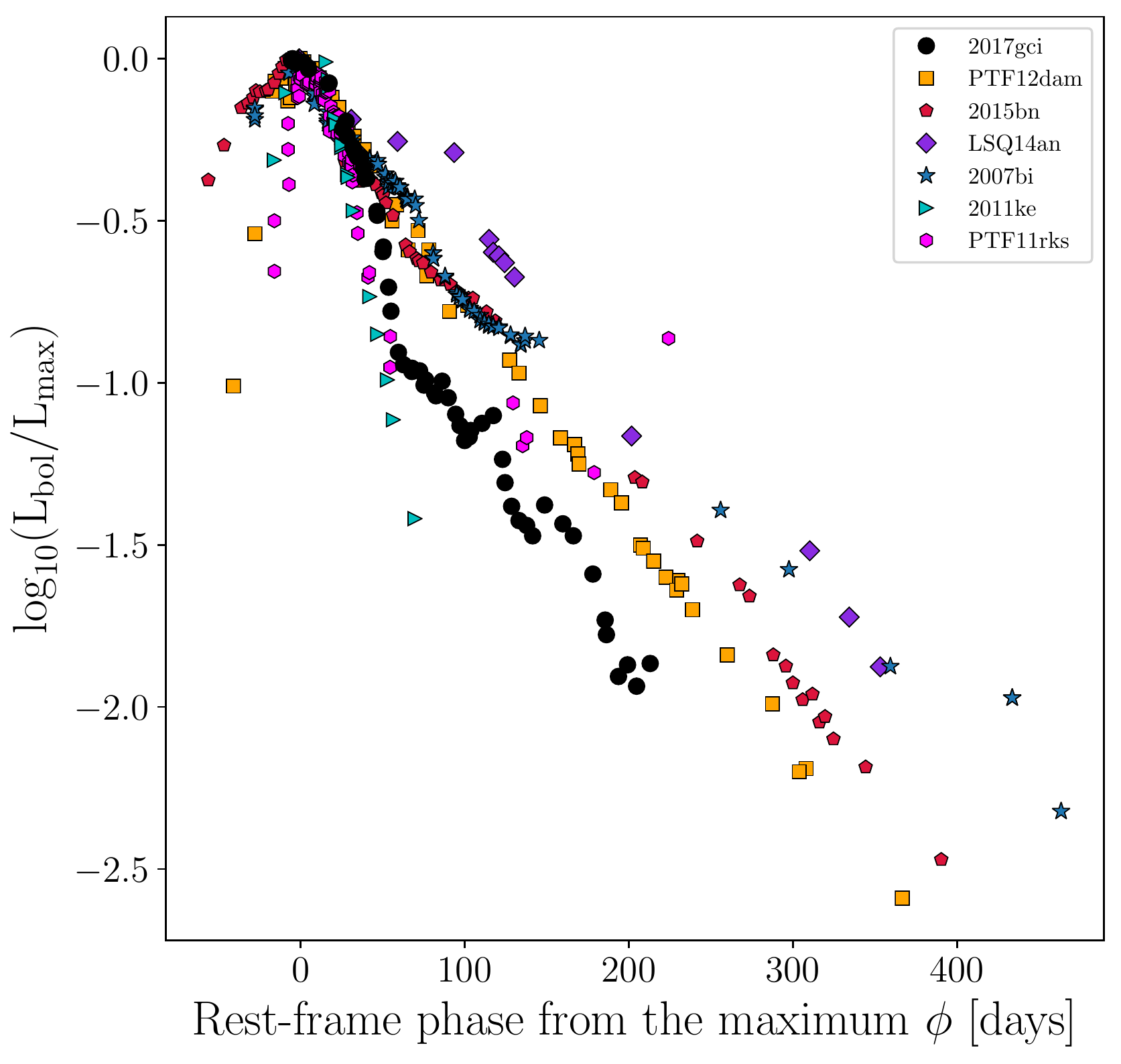}
\caption{Comparison of the bolometric LC of SN~2017gci (black dots) with those of SN 2015bn (red pentagons), PTF12dam (orange squares), SN 2007bi (purple stars), 2011ke (cyan triangles) and PTF11rks (magenta hexagons).}
\label{fig:comp}
\end{figure}
\begin{figure*}
\centering
\includegraphics[width=10cm]{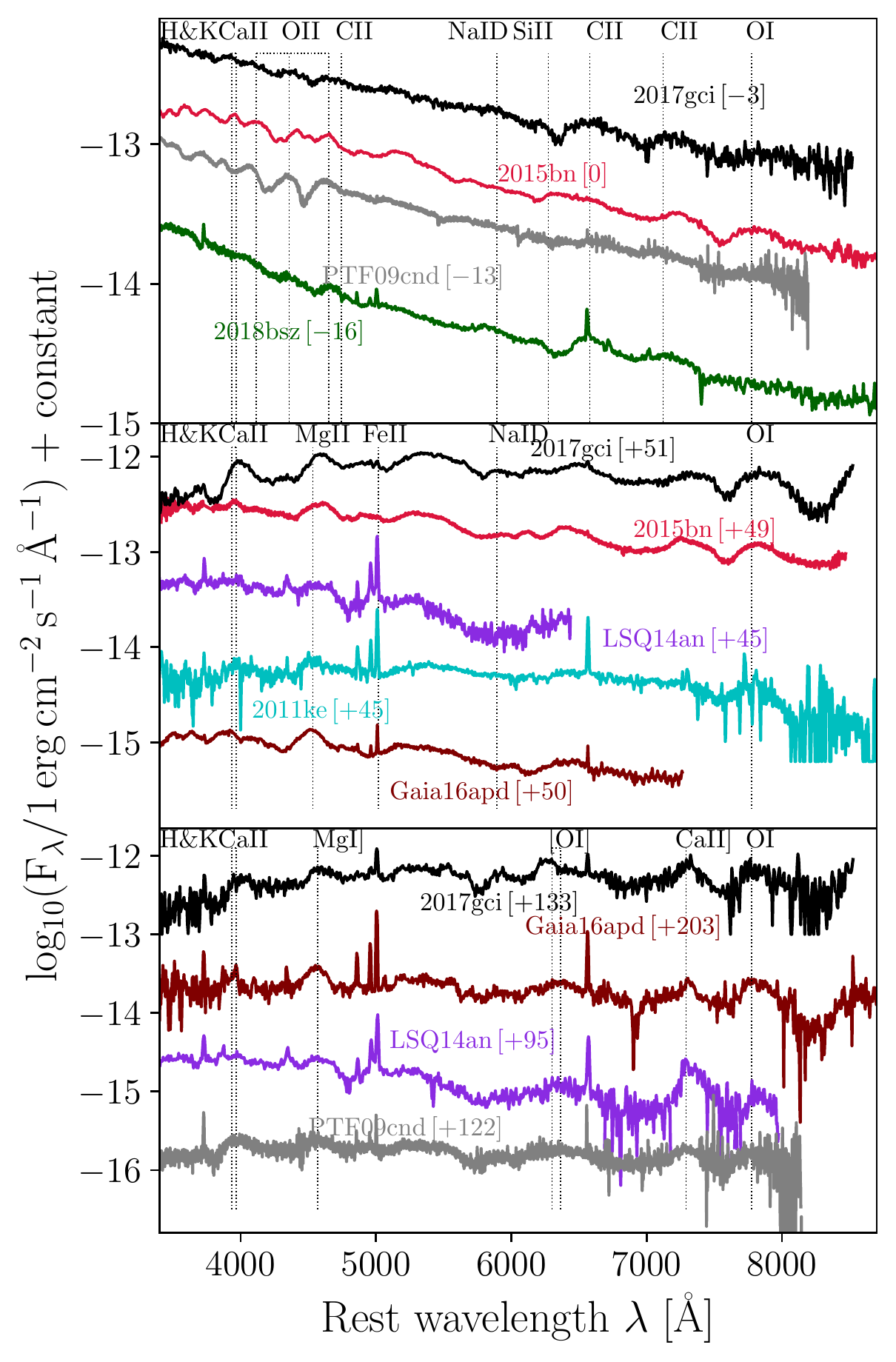}
\caption{Comparisons of three spectra of SN~2017gci (black line) with other SLSNe~I spectra at different phases with respect to the maximum luminosity epoch. Each spectrum is identified by the name of the SLSN~I and its rest-frame phase from the maximum light (in square brackets), and the labels are coloured as the spectrum they refer to. Line-identifications (black dotted lines) are also provided. Top panel: SLSN~I spectra at pre-maximum and maximum. Mid panel: SLSN~I spectra about ~40 days after the maximum. Bottom panel: SLSN~I spectra at late phases ($\gtrsim100$ days) after the maximum-luminosity epoch. The spectra of PTF09cnd {\citep{quimbyetal2018}}, SN 2018bsz {\citep{andersonetal2018}}, Gaia16apd {\citep{kangasetal2017}} and LSQ14an {\citep{inserraetal2017}} were taken from The Open Supernova Catalog. {For the spectra of SN 2011ke (mid panel) and of SN~2017gci at $\phi=133$ (bottom panel), the narrow emissions lines from the host galaxy were cut for display purposes.}}
\label{fig:speccomp}
\end{figure*}
\subsubsection{Photospheric velocity}
\label{sec:phrii}
To estimate the photospheric velocity we measured the wavelengths corresponding to the minima of the P-Cygni profiles which occur in the spectra of SN~2017gci. {They were determined with a gaussian fit of the absorption features (see Fig.~\ref{fig:vphot}) after having been normalized and continuum-subtracted.} We performed these measurements {from {$\phi=-7$ to $\phi=-4$} days, when the O {\scriptsize  II} absorption minima are present in the spectra. Errorbars are estimated by changing the continuum level multiple times before performing the fit}. The Doppler shift measured with respect to the rest-frame wavelength of the emissions corresponds to a photospheric velocity {$v(\mathrm{O\,{\scriptsize{II} ) }}\lesssim 8000\,\mathrm{km\,s^{-1}}$} (see Fig.~\ref{fig:vphot}). 
\begin{figure*}
\centering
\includegraphics[width=13 cm]{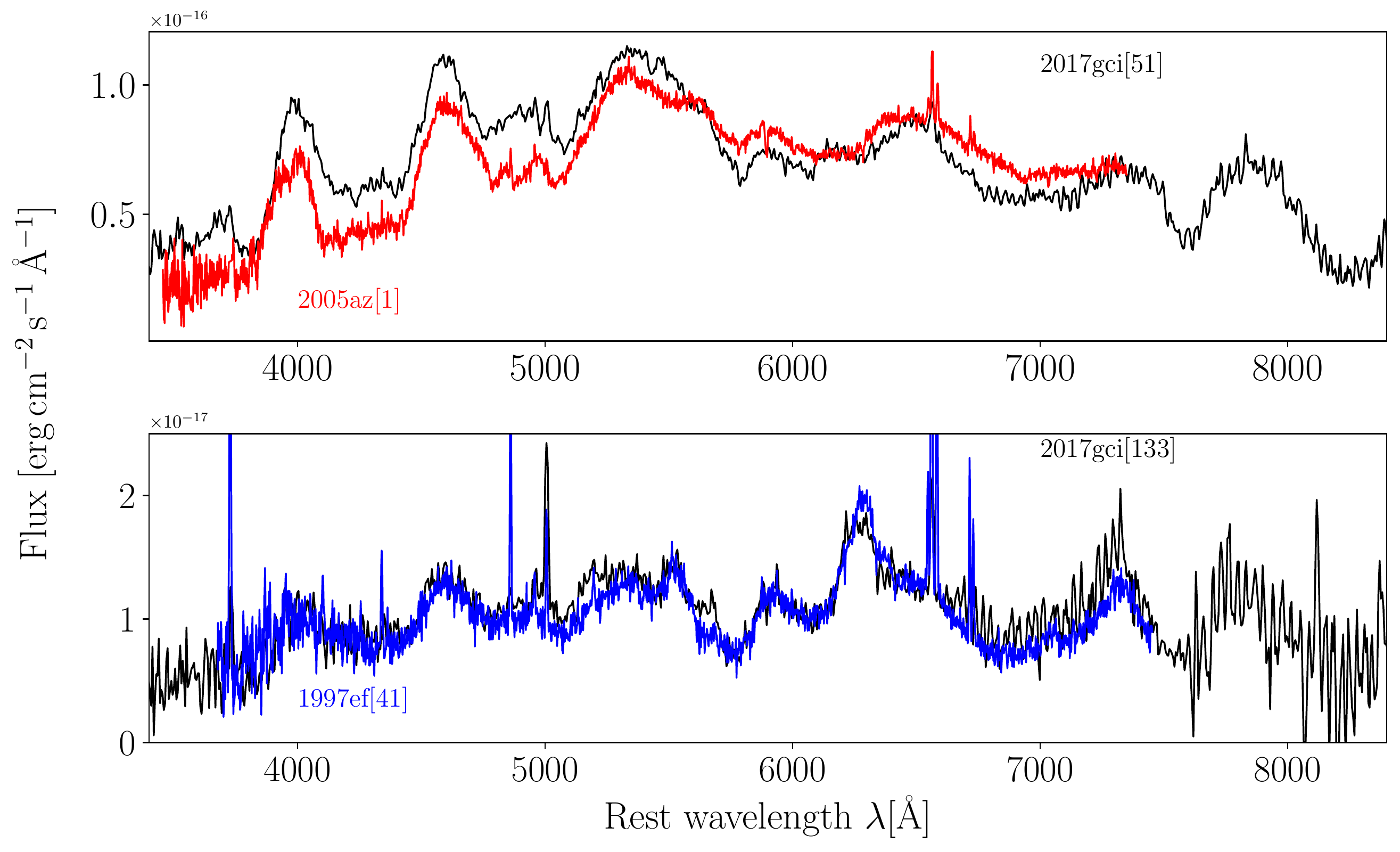}
\caption{Spectral comparisons of SN~2017gci obtained with the tool GELATO. The SN name is labelled in the plot alongside the rest-frame phase $\phi$ with respect to the maximum light (in square brackets). Top panel: SN~2017gci (black, {$\phi=51$} days) with the spectrum of the Type Ic SN 2005az (red, $\phi=1$ day). Bottom panel: SN~2017gci (black, {$\phi=135$} days) with the spectrum of the type Ic BL SN 1997ef (blue, $\phi=41$ day). }
\label{fig:gelato}
\end{figure*}
\section{Discussion}
In the following, we will discuss the interpretation of the data presented in the previous Sections.

\subsection{Metallicity of the host galaxy}
\label{sec:metall}
We estimated the metallicity of the SN~2017gci site by means of the narrow emission lines of the spectra at {$\phi=187,367$} days, attributed to the host-galaxy contribution. To test simultaneously several metallicity diagnostics, we used the python-based tool PYMCZ \citep{biancoetal2016}. PYMCZ takes as input a list of flux measurements with an associated uncertainty for [O {\scriptsize II}] $\lambda3727$, H$\beta$, [O {\scriptsize III}] $\lambda4959$, [O {\scriptsize III}] $\lambda5007$, H$\alpha$, [N {\scriptsize II}] $\lambda6584$, [S {\scriptsize II}] $\lambda6717$. For each flux measurement, PYMCZ generates a set of synthetic data via a Monte Carlo simulation. Hence a Gaussian probability distribution is drawn (whose mean is the input flux and whose standard deviation is the uncertainty of the flux) and randomly sampled. These flux measurements are used to compute the $12+\log_{10}(\mathrm{O/H})$ index via the D02 \citep{denicoloetal2002}, PP04 N2Ha, PP04 O3N2 \citep{pettiniandpagel2004}, M08 N2Ha, M08 O3O2 \citep{maiolinoetal2008} and M13 N2 \citep{marinoetal2013} calibrators. The resulting $12+\log_{10}(\mathrm{O/H})$ estimates (see the boxplot in Fig.~\ref{fig:pymcz}) cluster around $\sim8.1$ ($\sim0.3\,Z_\odot$), thus pointing towards a low-metallicity environment as it is expected by the host galaxies of SLSNe~I (see Introduction). A comparison of the PP04 O3O2 metallicity measurements of SN~2017gci with other SLSNe~I and GRBs at redshift $z\lesssim0.1$ is reported in Tab. \ref{tab:metall}. The environment of SN~2017gci is among those with the lower metallicity.

\subsection{[O {\scriptsize I}] emission profile}

In the spectrum of SN~2017gci at $\phi=187$ days, a close look to the profile of the [O {\scriptsize  I}] $\lambda\lambda\,6300,6364$ emission doublet points out the presence of a double peak on its topside (see Fig.~\ref{fig:oi636364}, top panel). The bluest hump of the [O {\scriptsize I}] doublet peaks at $\lambda\sim6260$ \AA{} (i. e. $\sim40$ \AA{} blueshifted with respect to its rest-frame wavelength) while the reddest hump peaks between $\lambda\sim6300-6310$ \AA{}. The $40-50$ \AA{} separation between the two peaks (which is lower than the natural 64 \AA{} separation of the doublet) is similar to what was found by \citet{milisavljevicetal2010} for the velocity shifts measured on the asymmetric [O {\scriptsize  I}] profiles of a sample of stripped-envelope SNe. 
In fact, double or multi-peaked [O {\scriptsize  I}] profiles were also observed in the late spectra of SNe Ib/c, as in the case of SN~2005bf \citep{anupamaetal2005} or SN~2009jf \citep[][]{valentietal2011,sahuetal2011} \citep[see also][for further studies on the asymmetric \textrm{[O {\scriptsize I}]} profiles]{taubenbergeretal2009}.

The $\sim40$ \AA{} wavelength shift of the blue peak corresponds to a velocity blueshift of $\sim2000\,\mathrm{km\,s^{-1}}$. To test whether the observed double-peaked [O {\scriptsize I}] profile could be reproduced by two velocity components, we fitted a composite model made of five gaussians (see Fig.~\ref{fig:oi636364}, lower panel): two gaussians for the $2000\,\mathrm{km\,s^{-1}}$-blueshifted [O {\scriptsize I}]-doublet component (with a FWHM of $\sim4000\,\mathrm{km\,s^{-1}}$), two gaussians for the rest-frame [O {\scriptsize I}]-doublet component (with a FWHM of $\sim4000\,\mathrm{km\,s^{-1}}$) and a broader ($\mathrm{FWHM}\sim11000\,\mathrm{km\,s^{-1}}$) rest-frame component. The FWHM of the two doublets was kept constant in the fitting procedure. The peaks of two couples of gaussians have a fixed separation of 64 \AA{} and a flux ratio 3:1 (see Fig.~\ref{fig:oi636364}, lower panel). The broad component was added to better fit the broad wings of the emission feature. The best-fit curve (see Fig.~\ref{fig:oi636364}, lower panel) underestimates the flux emitted in the blue hump of the doublet, but broadly accounts for the $40-50$ \AA{} separation of the two peaks.

The physical interpretation of the such profiles is not unique. It was suggested \citep{taubenbergeretal2009,valentietal2011} that they may be the signature of a certain degree of ejecta asphericity. In fact, an asymmetric jet-like explosion (where the major part of the material is launched in the direction opposite to the observer) or ejecta blobs could be responsible of the two velocity components. In addition, the wings could be explained by a more spherically-symmetric ejecta component.

Finally, further clues on the ejecta geometry of SN~2017gci will be given by polarimetric observations. In fact, continuum polarization measurements of SN~2017gci display an evolution in the polarization degree, which grows for $\phi>27$ days. This may be an evidence of the SN photosphere departure from spherical symmetry at late phases (Cikota et al., in preparation).

\subsection{Comparisons with other SLSNe~I}
\subsubsection{Comparing the bolometric light curves}
We compared the bolometric LC of SN~2017gci with those of a sample of SLSNe~I. Among these, the slow-SLSNe~I subsample consists of SN 2015bn \citep{nicholletal2016a,nicholletal2016b}, PTF12dam \citep{nicholletal2013,chenetal2015,vreeswijketal2017}, SN 2007bi \citep{galyametal2009}, {PTF09cnd \citep{quimbyetal2018}, SN 2018bsz \citep{andersonetal2018}} and LSQ14an \citep{inserraetal2017}, whereas the fast subsample includes SN 2011ke and PTF11rks \citep{inserraetal2013} (see Fig.~\ref{fig:comp}). The apparent magnitudes of the last two were taken from The Open Supernova Catalog \citep[\texttt{https://sne.space/},][]{guillochonetal2017}. Soon after the maximum luminosity, the LC decline of SN~2017gci is much faster than the slow SLSNe, except for SN 2015bn which shows an initial change of slope after the maximum luminosity. This might suggest that SN~2017gci is a fast SLSN~I, as confirmed by the comparison with SN 2011ke and PTF11rks which fairly well reproduce the decline of SN~2017gci. 
\subsubsection{Spectroscopic comparison}
Moreover, three spectra of SN~2017gci (at {$\phi=-7,51,133$} days) have been compared with the spectra of other SLSNe~I (see Fig.~\ref{fig:speccomp}). To the previous comparison sample, we added also two spectra of the intermediate-evolving type I SLSN Gaia16apd \citep{kangasetal2017}.
At pre-maximum/maximum epochs the spectral features of SN~2017gci show similarities with those of two slow SLSNe~I, namely SN 2015bn and SN 2018bsz \citep{andersonetal2018}. In particular, the presence of the broad C {\scriptsize  II} features on the red side of the spectrum makes SN~2017gci look like a slow SLSN~I. 

At later phases, the spectra become more similar each other. After $\sim40$ days from maximum light, the spectra show several broad features and nearly reproduce the overall spectral behaviour of SNe Ic BL at maximum luminosity. This actually holds true both for the slow and fast-evolving SLSNe~I (whose prototype is SN 2011ke). Similarly, for $\phi\gtrsim100$ days, the spectrum of SN~2017gci has characteristics similar to the other SLSNe of the sample, with the presence of Mg {\scriptsize  I}], [Ca {\scriptsize  II}] and the O {\scriptsize  I} emissions. As already {mentioned}, the resemblance of the late ($\phi\gg50$ days) post-maximum spectra of SN~2017gci with those of a SN Ic BL was verified via the GEneric cLAssification TOol \citep[GELATO,][]{harutyunyanetal2008} which, for the spectra at {$\phi=51,133$} days respectively outputs as best-match template SN 2005az (a Type Ic SN, $\phi=1$ day) and  SN 1997ef (a type Ic BL SN, at $\phi=41$ days in Fig.~\ref{fig:gelato}). In particular, the remarkable similarity of the spectrum of SN~2017gci with that one of SN 1997ef implies that they share the same chemical composition and kinematic.
\subsubsection{Temperature evolution}
\label{sec:tempevol}
Fig.~\ref{fig:bbrad} shows the temporal evolution of the blackbody temperature $T_\mathrm{BB}$ deduced by fitting the SED with a blackbody curve. {From the comparisons between SED and spectra we estimated that the maximum error introduced by deriving the BB temperature from the SED alone is about 1500 K.}
We see that at about 50 days after the maximum light the temporal evolution of $T_\mathrm{BB}$ settles on a plateau of $\sim4000-6000$ K. {This behaviour is similar to the flattening reached at $\sim6000-8000$ K by the evolution of the blackbody temperature of other SLSNe \citep[e. g. SN 2015bn and SN 2011ke, see Fig.~\ref{fig:bbrad} and Fig.~5 in][]{inserraetal2013}. \citet{nichollguillochonandberger2017} proposed that such a temperature floor might be due to O {\scriptsize II} recombination or to the instability-driven fragmentation of a dense shell. In the case of SN 2017gci, this happens around the same epoch of the knee ($\sim55-60$ days). Therefore we disfavour the possibility that the O {\scriptsize II} recombination might be responsible of this temperature levelling since in the O {\scriptsize II} $\lambda\lambda\,4115,4357,4650$ features already disappeared at $\phi=22$ days.}    

\subsection{Modelling the photometric evolution of SN~2017gci}
\label{sec:model}
In order to explain the photometric evolution of SN~2017gci, we considered {three} possible power sources: the $^{56}\mathrm{Ni}$ decay{, }the spin-down radiation from a central highly-magnetized neutron star (i.e. a newly-born magnetar), and the SN ejecta-CSM interaction. 
We explored the possible contribution from these different power sources under the assumption that the presence of the late LC bumps is to be attributed to the interaction of the SN ejecta with CSM shells or clumps (see below) and, following the approach illustrated in Section \ref{sec:mclc}, we excluded such bumps when fitting the different models.


Both the $^{56}\mathrm{Ni}$ decay- and the magnetar-powered synthetic LCs were computed via the semi-analytic diffusion scheme described in \citet[][hereafter referred to as I13]{inserraetal2013}, whose formalism was introduced by \citet{arnett1982}. This scheme relies upon three fundamental assumptions:
\begin{enumerate}
\item the input-power source is centrally located and the ejecta expand in a homologous, spherically symmetric way;
\item the opacity $\kappa$ is independent of time, density $\rho$ and temperature $T$;
\item radiation-dominated conditions of the environment, hence the radiation pressure $P_\mathrm{rad}$ is such that:
\begin{equation}
P\simeq P_\mathrm{rad}=\frac{1}{3}aT^4\,,
\end{equation}
where $P$ is the total pressure and $a=7.5646\times10^{15}\,\mathrm{erg\,cm^{-3}\,K^{-4}}$ is the radiation-density constant.
\end{enumerate} 
The instantaneous energy balance of such a physical system can be described by the first law of thermodynamics differentiated with respect to time:
\begin{equation}
\label{eq:flt}
\dot{U}-\frac{P}{\rho^2}\dot{\rho}=\epsilon-\frac{\partial L}{\partial m}\,,
\end{equation}
where $U$ is the internal energy per unit mass and the dot notation indicates the time derivative. The right-hand side of (\ref{eq:flt}) represents the heat-exchange variation, expressed as the sum of the specific input power $\epsilon$ and the luminosity radiated away per unit mass, $-\partial L/\partial m$. Here $L$ is expressed in diffusion approximation:
\begin{equation}
L=\frac{4\pi cr^2}{3\kappa\rho} \frac{\partial aT^4}{\partial r}\,,    
\end{equation}
where $c$ is the speed of light. Under these assumptions, a LC model is then obtained as a particular solution of (\ref{eq:flt}), substituting for $\epsilon$ either the power per unit mass from the $^{56}$Ni decay:
\begin{equation}
^{56}\mathrm{Ni}\rightarrow^{56}\mathrm{Co}\rightarrow^{56}\mathrm{Fe}\,,
\end{equation} or the magnetar spin-down luminosity divided by the ejecta mass.

Following the above prescription, we first performed a pure-{$^{56}$Ni-powered} fit and found a best-fit LC profile that cannot reproduce the data and, most importantly, a $^{56}\mathrm{Ni}$ mass even greater than the total ejecta mass. This is, as expected, a non-viable choice to explain the bolometric LC of SN~2017gci, therefore we excluded this possibility. 

\begin{figure*}
\centering
\includegraphics[width=10cm]{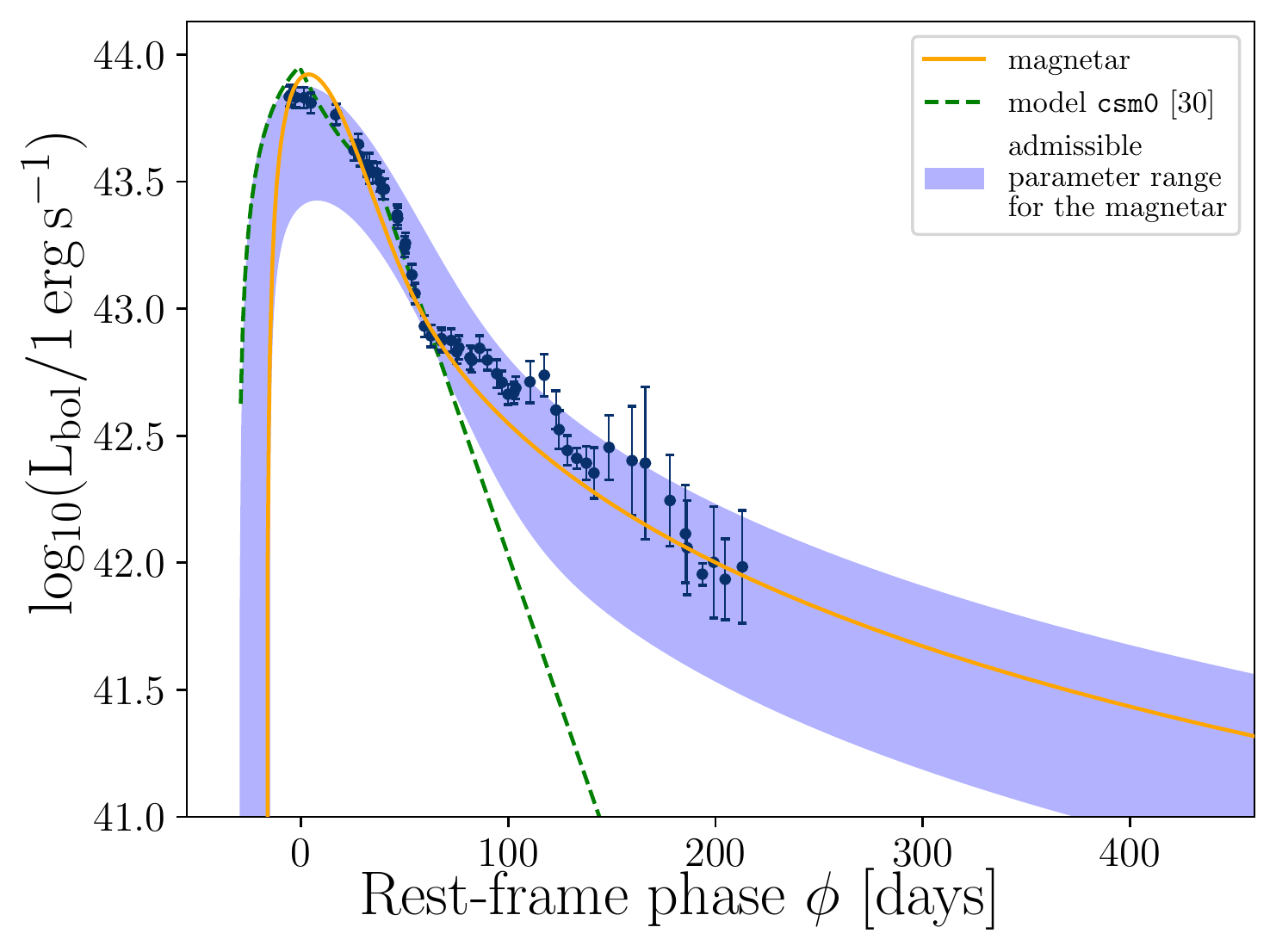}
\caption{Pseudo-bolometric LC of SN~2017gci {compared }with a magnetar-powered LC fit {(yellow line), a constant-density CSM shell model obtained with \texttt{TigerFit} (green dashed line) assuming a phase from the explosion $\phi_0$ of 30 days. 
In addition, the blue-shaded area covers the admissible parameter range for a magnetar model fit}. }
\label{fig:mult}
\end{figure*}

Then, we considered a magnetar power source (Fig.~\ref{fig:mult}), with the physical parameters of the fit being the ejecta mass $M_\mathrm{ejecta}$, the polar magnetic field $B_\mathrm{p}$, the initial orbital period of the magnetar $P_\mathrm{initial}$, the phase from the maximum luminosity epoch $\phi_0$, and the effective opacity $\kappa$.
In Tab.~\ref{tab:fp} we report the best-fitting magnetar model parameters (``MF1''), including a rise time of {$\approx16$} rest-frame days, consistent with the fast-evolving interpretation \citep{inserra2019}.
\begin{figure*}
\centering
\includegraphics[width=16 cm]{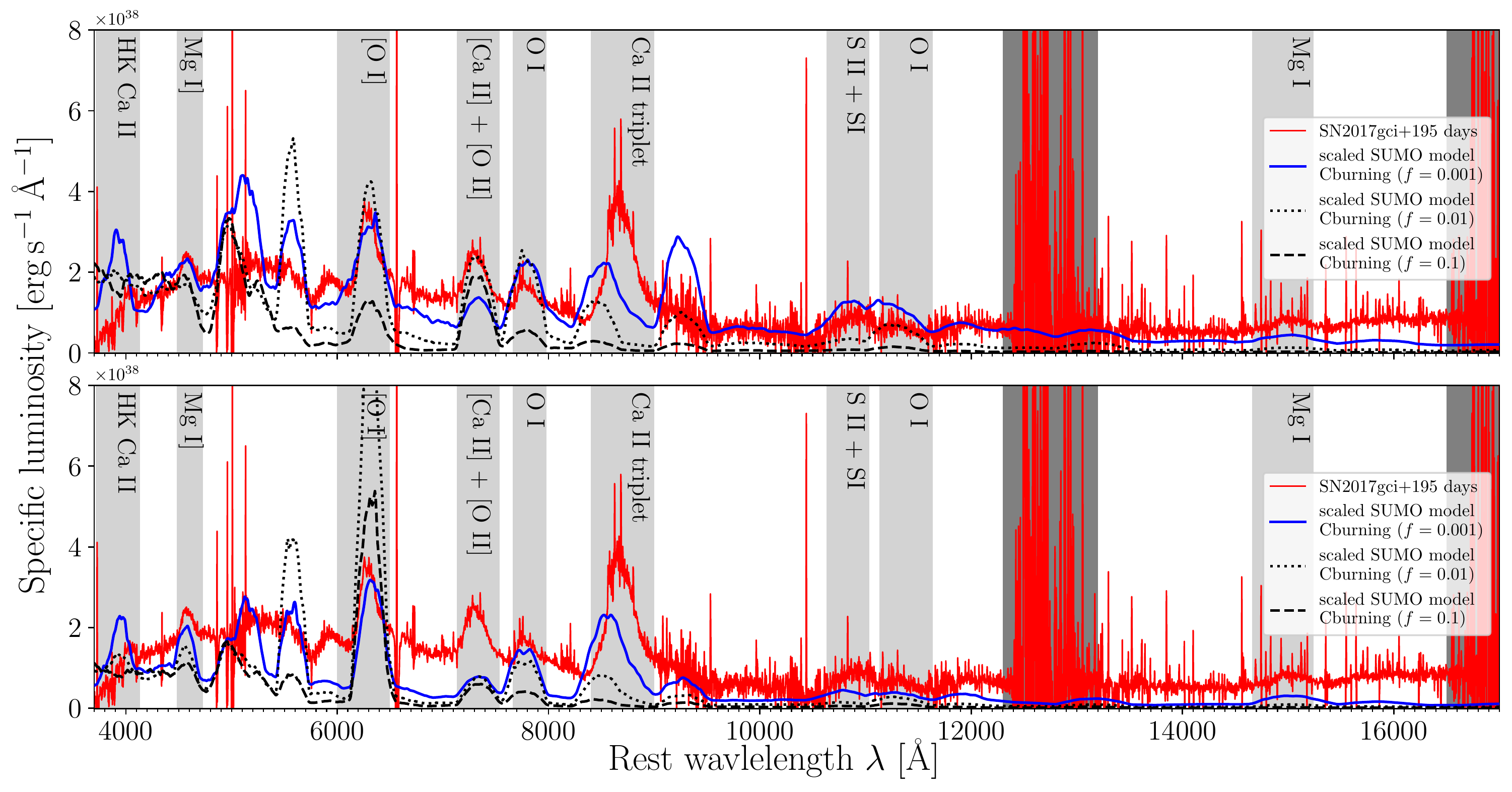}
\caption{Comparison of the spectrum of SN~2017gci {at {$\phi=187$ days}} (red solid line) with the models output by the SUMO code \citep[][see text]{jerkstrandetal2017}. A residual host contribution was estimated and subtracted to the original one (see text). Upper panel: models with $M_\mathrm{ejecta}=10\,\mathrm{M_\odot}$, energy deposition $E_\mathrm{dep}=2\times10^{42}\,\mathrm{erg\,s^{-1}}$ and $f=0.001$ (blue solid line), 0.01 (black dotted line) and 0.1 (black dashed line). Lower panel: as before, but with models whose energy deposition is $E_\mathrm{dep}=5\times10^{41}\,\mathrm{erg\,s^{-1}}$. These were scaled by a factor 2. The light-gray shaded area mark the features of the spectrum, whereas the dark-gray ones are placed in correspondence with the atmospheric corrections of the XS-reduction pipeline \citep{kruehleretal2014}.}
\label{fig:sumo}
\end{figure*}
{While the best fit magnetar model is able to describe the maximum luminosity of SN~2017gci, it does not entirely explain the behaviour of the observed bolometric LC at later times. In particular, between about 100 and 200 days, the observed luminosity is significantly higher and requires an additional power source. As discussed below, this could be the interaction of the ejecta with CSM clumps. 

The above result depends on the exclusion from the fitting procedure of data points in two different intervals between about 100 and 200 days (see Fig.~\ref{fig:blc}). In order to provide also a more conservative indication, we explored in full the admissible parameter range for a magnetar LC in presence of all data points (shaded region in Fig.~\ref{fig:mult}). 
We found that the observed bolometric LC could not be easily described by the magnetar scenario solely outside the following ranges: $3.5\times10^{14}\,\mathrm{G}\lesssim B_\mathrm{p}\lesssim8\times10^{14}\,\mathrm{G}$, $1\,\mathrm{ms}\lesssim P_\mathrm{initial}\lesssim5\,\mathrm{ms}$, $7.7\,\mathrm{M_\odot}\lesssim M_\mathrm{ejecta}\lesssim12\,\mathrm{M_\odot}$, $0.08\,\mathrm{cm^2\,g^{-1}}\lesssim\kappa\lesssim0.2\,\mathrm{cm^2\,g^{-1}}$, and $18\,\mathrm{days}\lesssim\phi_0\lesssim30\,\mathrm{days}$.


Finally, we considered the SN ejecta-CSM interaction as the main power source for the bolometric LC, in particular at maximum luminosity.
To fit such a model, we used \texttt{TigerFit}.
Since \texttt{TigerFit} works at fixed phase from the explosion $\phi_0$, we assumed different values for it between 18 and 30 days. \texttt{TigerFit} embeds the modules \texttt{csm0} and \texttt{csm2}, which refer to the case of a steady-state wind 
and a constant-density CSM shell respectively.\footnote{{The two modules are labelled with the value of the exponent $s$ of the density-profile slope $\rho_\mathrm{CSM}\propto r^{-s}$. Hence $s=2$ corresponds to a wind solution and $s=0$ to a constant-density shell. }}
Using both modules and fitting only up to the knee (at about {54-57} days), 
we found the best agreement with the \texttt{csm0} model for $\phi_0=30$ days. This gives a CSM mass of 4.9 $\mathrm{M}_\odot$, for a progenitor radius $R_\mathrm{progenitor}\simeq0.004\times10^{14}\,\mathrm{cm}$ (see Tab.~\ref{tab:fp} for further details). As shown in Fig.~\ref{fig:mult} (green dashed profile, referring to $\phi_0=30$\,days), this model is able to reproduce the maximum luminosity of SN~2017gci, although any data point after the knee require an additional power source. 
Overall, we conclude that the maximum and initial part of the LC could be explained with a magnetar power source or with the ejecta-CSM interaction.}

As mentioned above, we assumed that the luminosity undulations observed between the knee and {$\phi\simeq200$} days can be explained via the interaction of the ejecta with CSM clumps. This appears to be the most natural explanation of the bumps in the LCs of SLSNe~I \citep{moriyaetal2018}. 
Following \citet{nicholletal2016a}, we attempted an estimate of the mass of the CSM clumps $M_\mathrm{CSM}$ based on the simple relation 
\begin{equation}
E_\mathrm{rad}\simeq\frac{1}{2}M_\mathrm{CSM}u_\mathrm{ej-CSM}^2\,,
\end{equation}
where $E_\mathrm{rad}$ is the energy radiated at the epochs of the LC bumps and $u_\mathrm{ej-CSM}$ is the relative velocity between the ejecta and the CSM clumps. 
Taking our best fitting magnetar model as a reference, $E_\mathrm{rad}$ was computed by integrating the difference between the bolometric LC and the magnetar fit between $\phi=101$ and $\phi=130$ days for the first bump and between $\phi=138$ and $\phi=196$ days for the second one. We obtained $E_\mathrm{rad,1}\simeq3.5\times10^{48}$ erg and $E_\mathrm{rad,2}\simeq2.8\times10^{48}$ erg. Then, assuming that the CSM has a negligible velocity compared to the ejecta, we have
\begin{equation}
M_\mathrm{CSM,1}\simeq5.5\times10^{-3}\left(\frac{u_\mathrm{ejecta}}{8000\,\mathrm{km\,s^{-1}}}\right)^{-2}\,\mathrm{M_\odot}
\end{equation}
and
\begin{equation}
M_\mathrm{CSM,2}\simeq4.4\times10^{-3}\left(\frac{u_\mathrm{ejecta}}{8000\,\mathrm{km\,s^{-1}}}\right)^{-2}\,\mathrm{M_\odot}\,.  
\end{equation}

The interaction postulated between the SN ejecta and the CSM clumps may possibly leave its signature in the optical spectra (like the presence of narrow emission lines). However, the spectrum at phase {$\phi=103$} days (nearly corresponding to a bump) does not show any sudden difference compared to the subsequent one (at {$\phi=133$} days, which is about a minimum of the LC undulations). This can be understood also from the undulations in Fig.~\ref{fig:undul}, which do not show a noticeable wavelength dependence in optical bands, suggesting that in this case the hypothetical CSM interaction had a grey effect on the opacity of SN~2017gci. 
\begin{figure}
\centering
\includegraphics[width=5 cm]{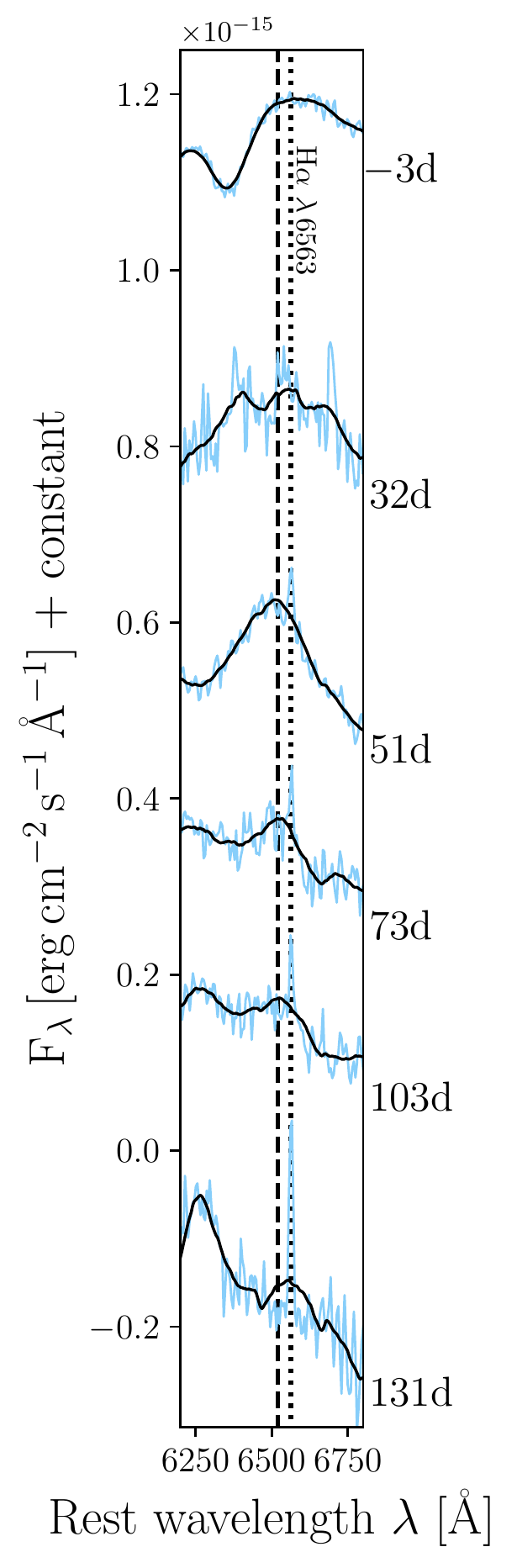}
\caption{{The broad feature around $\lambda\simeq6520$ \AA{} emerging at {$\phi=51$} days and  broadly consistent with H$\alpha$. The black dashed line indicates the centroid of the emission and the black dotted lines marks the rest wavelength of the H$\alpha$ line. Similarly to Fig.~\ref{fig:spec_evol}, the light-blue solid line is the observed spectrum after having subtracted a blackbody continuum. The black solid line is the same but smoothed with a Savitzki-Golay filter. The rest-frame phase is labelled on the right for each continuum-subtracted spectrum.}}
\label{fig:halpha}
\end{figure}

{In addition, if the spectral feature at about $\lambda\sim6520$ \AA{} {(see Section \ref{sec:spectra})} is indeed $\mathrm{H}\alpha$, it could be a signature that the interaction with a hydrogen-rich CSM  has started {already before} the knee: this could account for the almost linear decline of the LC (see Section \ref{sec:spec}), thus supporting the idea that the bumps are caused by overdensities in the CSM.  This is also supported by the fact that the H$\alpha$-like feature apparently disappears from the spectrum at {$\phi=133$} days which is consistent with the disappearing of the first bump (see Fig.~\ref{fig:halpha}). Hence the second bump could be the signature of a hydrogen-poor clump of CSM, although this would require the presence of CSM clumps with very different chemistry. }
\subsection{SN~2017gci at nebular phases}
\label{sec:nebularph}
We compared the spectrum of SN~2017gci at {$\phi=187$} days thanks to spectral-modelling numerical code SUMO \citep{jerkstrandetal2011,jerkstrandetal2012,jerkstrandetal2017}. The publicy-available\footnote{\texttt{https://star.pst.qub.ac.uk/wiki/doku.php/users/
\\
ajerkstrand/start/} .} SUMO models for SLSNe I \citep{jerkstrandetal2017} are computed at 400 days after the explosion, at a constant ejecta velocity $v_\mathrm{ejecta}=8000\,\mathrm{km\,s^{-1}}$ and with $N=100$ random clumps for different ejecta compositions \footnote{Pure Oxygen, C-burning ashes, Oxygen (92\%) and Magnesium (8\%).}, ejecta masses $M_\mathrm{ejecta}$, energy deposition $E_\mathrm{dep}$ and filling factors\footnote{The filling factor $f$ expresses the percentage volume of clumps. Hence $1-f$ corresponds to vacuum.} $f$. Before adapting a SUMO solution to the observed spectrum at {$\phi=187$} days, we estimated the residual contribution of the host galaxy emission therein, similarly to \citet[][see their Section 2]{jerkstrandetal2017}. Since a spectrum of the host galaxy is not available, we took a starbust-galaxy template spectrum from the sample of \citet{calzettietal1994}, which was obtained by averaging over a sample of starbust galaxies with $0.11<E(B-V)<0.21$ mag. Then we scaled the template spectrum on the SED of the host galaxy, which we measured from the template images (see Section \ref{sec:phot}). Hence we subtracted the scaled template spectrum to the XS spectrum at {$\phi=187$} days (see Fig.~\ref{fig:sumo}).  We found that the [O {\scriptsize  I}] $\lambda\lambda\,6300,6364$ is well reproduced by the spectral models computed with C-burning composition (O/Mg/Ne dominated) with $M_\mathrm{ejecta}=10\, \mathrm{M}_\odot$, $N_\mathrm{clumps}=100$, $f=0.001$ with an energy deposition $E_\mathrm{dep}=5\times10^{41}\,\mathrm{erg\,s^{-1}}$ or  $2\times10^{42}\,\mathrm{erg\,s^{-1}}$ (see Fig.~\ref{fig:sumo}). The latter were scaled by a factor 2 (see the caption of Fig.~\ref{fig:sumo}) to roughly fit the luminosity of the spectrum.
The ejecta-mass value of the two models is fairly similar to the one obtained from the fit of the magnetar-powered synthetic LC (see Tab.~\ref{tab:fp}) as well as the magnetar energy deposition, which in MF1 is $\sim9.44\times10^{41}\,\mathrm{erg\,s^{-1}}$ for {$\phi=187$} days.

Moreover, for both the two models we showed the effect of increasing the filling factor. The luminosities of the Mg {\scriptsize  I}] $\lambda4571$ and Mg {\scriptsize  I} $\lambda15400$ suggest that the ejecta are likely clumped with a filling factor $f\ll1$ \citep{jerkstrandetal2017}. This would not be surprising {in} {both} the magnetar {and in the CSM-interaction} scenario, where the SN ejecta are piled up by the pulsar-wind bubble in a high-density layer which is afterwards broken up by hydro-dynamical instabilities. However, if such a thin dense shell survived \citep[like in one-dimensional simulations, see e~g. Fig.~1 in ][]{kasenandbildsten2010}, it would result in a clear observational signature like, e~g., boxy-shaped spectral lines \citep[see][]{wheeleretal1990}  which we do not observe in the case of SN~2017gci. Recent two-dimensional hydrodynamic simulations of a magnetar-driven SLSN explosion \citep{chenetal2016} predict indeed that the low-density bubble inflated by the magnetar spin-down radiation becomes unstable to the onset of Rayleigh-Taylor Instabilities after the collision with the high-density SN ejecta. From a physical point of view, increasing the ejecta clumpiness enhances the effect of trace elements on temperature and ionization. In such a regime, nebular spectra are then more sensitive to the progenitor metallicity. Hence, the features shared by the nebular spectra of SLSNe I with those of Type Ic BL SNe might suggest that they have similar progenitors and/or explosion mechanisms \citep{nicholletal2016b}. 

Finally, it is possible to infer a physical-parameter estimate by means of the analytic relations discussed by \citet{jerkstrandetal2017}. Substituting the luminosity of the Oxygen recombination line luminosity $L_\mathrm{7774}$ (for which we measured $L_{7774}=1.1\times10^{40}\,\mathrm{erg\,s^{-1}\,}$), we constrained the electron density $n_e$ as follows:
\begin{equation}
\begin{split}
n_ef^{1/2}=4.20\times10^{7}\left(\frac{v_\mathrm{exp}}{8000\,\mathrm{km\,s^{-1}}}\right)^{-3/2}\left(\frac{\alpha^\mathrm{eff}(T)}{2\times10^{-13}\,\mathrm{cm^3\,s^{-1}}}\right)^{-1/2}\\
\,\mathrm{cm^{-3}},
\end{split}
\label{eq:o7774ne}
\end{equation}
where $\alpha^\mathrm{eff}(T)$ is the effective recombination rate and $v_\mathrm{exp}$ is the maximum velocity of the expanding gas, once spherical symmetry is assumed. The value that we obtained for $n_ef^{1/2}$ is in the range outlined by \citet{nicholletal2019} for a sample of 41 spectra of 12 SLSNe (see their Fig.~22). Then, assuming a filling factor $f=0.001${, a rise time of 16 days (as suggested by MF1),} $\alpha^\mathrm{eff}(T)=2\times10^{-13}\,\mathrm{cm^3\,s^{-1}}$ and $v_\mathrm{exp}=8000\,\mathrm{km\,s^{-1}}$, equation (\ref{eq:o7774ne}) implies $n_e\simeq1.3\times10^9\,\mathrm{cm^{-3}}$. This value agrees with the electron density that can be deduced from the intensity ratio of the  Ca {\scriptsize  II} NIR triplet and the forbidden  [Ca {\scriptsize  II}]. Such a ratio is particularly high in the case of the spectrum of SN~2017gci ({at} {$\phi=187$} days) since it reaches $\sim2.9$. This implies $n_e\gtrsim1\times10^{9}\,\mathrm{cm^{-3}}$ \citep[for a temperature of $6000-6500$ K, see Fig.~14 in ][]{jerkstrandetal2017}. This result is robust with respect to possible O line contaminations to the $\lambda$ 7300 and $\lambda$ 8600 features \citep[see][for details]{jerkstrandetal2017}. Such a high density at nebular phases could be achieved in principle by the matter swept-up by the pulsar-wind nebula. Finally, equation (9) from \citet{jerkstrandetal2017} provides an estimate of the magnesium mass $M_\mathrm{Mg}$ via the luminosity emitted within the Mg {\scriptsize I} 1.5 $\mu$m feature,  $L_{1.5\mu\mathrm{ m}}\simeq6.4\times10^{39}\,\mathrm{erg\,s^{-1}}$ with the following formula:
\begin{equation}
\label{eq:anders}
\frac{M_\mathrm{Mg}}{1\,\mathrm{M_\odot}}=\frac{L_{1.5\mu \mathrm{m}}}{6.6\times10^{38}\,\mathrm{erg\,s^{-1}}}\left(\frac{n_e}{10^8\,\mathrm{cm}^{-3}}\right)^{-1} \left(\frac{\alpha^\mathrm{eff}(T)}{1\times10^{-13}\,\mathrm{cm^3\,s^{-1}}}\right)^{-1}\,.
\end{equation}
Using the same value of $n_e$ and with $\alpha^\mathrm{eff}(T)=1\times10^{-13}\,\mathrm{cm^{3}\,s^{-1}}$, it gives $M_\mathrm{Mg}\simeq1\,\mathrm{M_\odot}$. Provided that the Mg mass fraction is typically 5-10\% of the O/Mg mass, the result of the equation (\ref{eq:anders}) implies a O/Mg zone mass $\lesssim10\,\mathrm{M_\odot}$. This also supports the choice of the $M_\mathrm{ejecta}=10\,\mathrm{M}_\odot$ models for SN~2017gci, given that only such a zone mass is consistent with the Mg {\scriptsize I} 1.5$\mu$m constraint. {These results might be in favour of the picture of a quite massive progenitor star \citep[$\gtrsim40\,M_{\odot}$, ][]{jerkstrandetal2017} for SN 2017gci.}
\section{Conclusions}
We have presented the optical/NIR photometry and the optical spectra of the nearby SN~2017gci, whose K-corrected absolute magnitude at maximum luminosity in $g$ band is $\sim-21.5$ mag. Its LC presents a sudden change in the slope {(the `knee')} and two bumps at $\sim110$ and $\sim160$ days after the maximum luminosity. Similar characteristics are not infrequent among the known slow SLSNe~I. Its spectroscopic evolution follows the typical one of SLSNe~I, which at about 40 days from the maximum light  turns into a SN-Ic BL-like spectrum at its maximum luminosity. 

We employed a semi-analytical model to fit the bolometric LC, assuming the following power sources: (i) the $^{56}\mathrm{Ni}$ decay chain; (ii) the magnetar-spin-down radiation; {(iii) the ejecta-CSM interaction}. The magnetar fit allowed for a physical parameter estimate which envisages an ejecta mass of about $\sim9\,\mathrm{M}_\odot$. This value is similar to those obtained for the SNe Ic BL \citep[as in the case of SN 1997ef,][]{nakamuraetal2000}. {Also, we performed a fit with the tool \texttt{Tigerfit} assuming that the CSM interaction contributes to the maximum luminosity. This requires a CSM mass of $\sim5\,\mathrm{M_{\odot}}$ and an ejecta mass of $\sim\,12\mathrm{M_{\odot}}$}. {In addition, we ascribe the presence of the knee and the bumps to the CSM interaction, which is supported by a likely presence of an {almost} coeval H$\alpha$ emission in the spectrum.} 

Additional indications were obtained from the moderate-resolution XS spectrum at phases {$\phi=187$} days thanks to a handful of spectral models produced via the SUMO single-zone code. We found {the best} agreement with models assuming $M_\mathrm{ejecta}=10\,\mathrm{M_\odot}$ and an energy deposition $E_\mathrm{dep}=5\times10^{41}-2\times10^{42}\,\mathrm{erg\,s^{-1}}$. Interestingly, {this broadly agrees with} the magnetar luminosity of MF1 at the rest-frame phase {$\phi=187$} days is $9.44\times10^{41}\,\mathrm{erg\,s^{-1}}$. \\
{Overall, our analysis points towards a progenitor mass of $\gtrsim40\,\mathrm{M_\odot}$ for SN~2017gci.}

The spectroscopic similarities between SLSNe~I and SNe Ic BL (e.~g., SN 1997ef) support the hypothesis that these SN subclasses are linked by a continuum distribution and share a similar origin {\citep{liuetal2017,deciaetal2018,quimbyetal2018,galyam2018b}}. However, the solution of the {SLSN-SN Ic BL} puzzle requires both a wider data sample and a further improvement of the modelling tools. In particular, we expect that next generation surveys such as the Legacy Survey of Space and Time (LSST) at the Vera Rubin Observatory (VRO) will discover a huge number of SLSNe \citep{villaretal2018}, which would be crucial especially for very early detections. In addition, three-dimensional hydrodynamical modelling including an improved treatment of radiative transport will allow us to better investigate the properties of SLSNe at both early and late phases, thus boosting our understanding of the underlying explosion mechanism \citep{sokerandgilkis2017} as well as the nature of the progenitor stars.
\section*{Data Availability Statement}
The data presented in this article and listed in the Appendix A are available in the online supplementary material.
\section*{Acknowledgements}
This article has been accepted for publication in MNRAS published by Oxford University Press on behalf of the Royal Astronomical Society.
{We thank the anonymous referee for the very useful comments, which contributed to improve the manuscript.}
 AF is partially supported by the PRIN-INAF 2017 with the project \textit{Towards the SKA and CTA era: discovery, localisation, and physics of transients sources} (P.I. M. Giroletti). {These observations made use of the LCO network.  DAH, CP, DH, and JB are supported by NSF Grant AST-1911225 and NASA Grant 80NSSC19k1639.} TMB was funded by the CONICYT PFCHA / DOCTORADOBECAS CHILE/2017-72180113. MG is supported by the Polish NCN MAESTRO grant 2014/14/A/ST9/00121. TWC acknowledges the funding provided by the Alexander von Humboldt Foundation and the EU Funding under Marie Sk\l{}odowska-Curie grant agreement No 842471, and Thomas Kr{\"u}hler for reducing X-Shooter spectrum. LG was funded by the European Union's Horizon 2020 research and innovation programme under the Marie Sk\l{}odowska-Curie grant agreement No. 839090. This work has been partially supported by the Spanish grant PGC2018-095317-B-C21 within the European Funds for Regional Development (FEDER). CPG acknowledges support from EU/FP7-ERC grant no. [615929]. GL was supported by a research grant (19054) from VILLUM FONDEN. MN is supported by a Royal Astronomical Society Research Fellowship. {R.L. is supported by a Marie Sk\l{}odowska-Curie Individual Fellowship within the Horizon 2020 European Union (EU) Framework Programme for Research and Innovation (H2020-MSCA-IF-2017-794467).} GT acknowledges partial support by the National Science Foundation under Award No. AST-1909796. Research by S.V. is supported by NSF grants AST-1813176  and AST-2008108. Some of the observations reported here were obtained at the MMT Observatory, a joint facility of the University of Arizona and the Smithsonian Institution under program 2018A-UAO-G16 (PI Terreran). Some of the data presented herein were obtained at the W. M. Keck Observatory, which is operated as a scientific partnership among the California Institute of Technology, the University of California, and the National Aeronautics and Space Administration under program NW440 (PI Fong). The Observatory was made possible by the generous financial support of the W. M. Keck Foundation. The authors wish to recognize and acknowledge the very significant cultural role and reverence that the summit of Maunakea has always had within the indigenous Hawaiian community. We are most fortunate to have the opportunity to conduct observations from this mountain. W. M. Keck Observatory and MMT Observatory access was supported by Northwestern University and the Center for Interdisciplinary Exploration and Research in Astrophysics (CIERA). Based on observations collected at the European Organisation for Astronomical Research in the Southern Hemisphere under ESO programmes 199.D-0143, 0100.D-0751(B), 0101.D-0199(B), 099.A-9025(A), 0100.A-9099(A)099.A-9099 and 0100.A-9099. This work makes use of observations from the LCO network. Part of the funding for GROND (both hardware as well as personnel) was generously granted from the Leibniz-Prize to Prof. G. Hasinger (DFG grant HA 1850/28-1). The Pan-STARRS1 Surveys (PS1) have been made possible through contributions of the Institute for Astronomy, the University of Hawaii, the Pan-STARRS Project Office, the Max-Planck Society and its participating institutes, the Max Planck Institute for Astronomy, Heidelberg, and the Max Planck Institute for Extraterrestrial Physics, Garching, The Johns Hopkins University, Durham University, the University of Edinburgh, Queen's University Belfast, the Harvard-Smithsonian Center for Astrophysics, the Las Cumbres Observatory Global Telescope Network Incorporated, the National Central University of Taiwan, the Space Telescope Science Institute, the National Aeronautics and Space Administration Grants No.s NNX08AR22G, NNX12AR65G, and NNX14AM74G, the National Science Foundation under Grant No. AST-1238877, the University of Maryland, Eotvos Lorand University (ELTE), the Los Alamos National Laboratory and the Gordon and Betty Moore foundation. The ATLAS surveys are funded through NASA grants NNX12AR55G. This work has made use of data from the European Space Agency (ESA)
mission {\it Gaia} (\url{https://www.cosmos.esa.int/gaia}), processed by
the {\it Gaia} Data Processing and Analysis Consortium (DPAC,
\url{https://www.cosmos.esa.int/web/gaia/dpac/consortium}). Funding
for the DPAC has been provided by national institutions, in particular
the institutions participating in the {\it Gaia} Multilateral Agreement. {This research made use of TARDIS, a community-developed software
package for spectral synthesis in supernovae
\citep{kerzendorfandsim2014}.
The development of TARDIS received support from the
Google Summer of Code initiative
and from ESA's Summer of Code in Space program. TARDIS makes
extensive use of Astropy and PyNE.}

\appendix
\section{Tables}
\begin{table*}
\caption{$UVW1,UVM2,UVW2$-filters observed (non K-corrected) {aperture} magnitudes (in AB system). Errors are in parentheses.}
\label{tab:uvottab}
\begin{tabular}{cccccc}
\hline
MJD&r. f. phase from maximum&$UVW1$&$UVM2$&$UVW2$&instrument\\
\hline
57986.00&{-3.96}&-&-&19.58(0.08)&{\textit{Swift}/UVOT}\\
57986.77&{-3.25}&-&19.46(0.05)&-&{\textit{Swift}/UVOT}\\
57989.82&{-0.44}&18.84(0.07)&-&-&{\textit{Swift}/UVOT}\\
57989.82&{-0.44}&-&-&20.04(0.10)&{\textit{Swift}/UVOT}\\
57990.82&{-0.44}&-&19.84(0.08)&-&{\textit{Swift}/UVOT}\\
57992.79&{2.29}&19.32(0.52)&-&-&{\textit{Swift}/UVOT}\\
57992.79&{2.29}&-&20.11(0.32)&-&{\textit{Swift}/UVOT}\\
57998.29&{7.35}&19.22(0.13)&-&-&{\textit{Swift}/UVOT}\\
57998.29&{7.35}&-&20.29(0.20)&-&{\textit{Swift}/UVOT}\\
57998.29&{7.35}&-&-&20.43(0.19)&{\textit{Swift}/UVOT}\\
58001.47&{10.28}&19.56(0.10)&-&-&{\textit{Swift}/UVOT}\\
58001.47&{10.28}&-&-&20.54(0.13)&{\textit{Swift}/UVOT}\\
58001.48&{10.28}&-&20.77(0.15)&-&{\textit{Swift}/UVOT}\\
\hline
\end{tabular}
\end{table*}
\begin{table*}
\caption{$g,r,i,z$-filter observed (non K-corrected, {non S-corrected}) magnitudes (in AB system). Errors are in parentheses.}
\label{tab:griztab}
\begin{tabular}{ccccccc}
\hline
MJD&r. f. phase from maximum&$g$&$r$&$i$&$z$&instrument\\
\hline
57931.00&{-54.55}&{21.10}({0.20})&-&-&-&{Gaia}\\
57977.44&{-11.83}&{17.30}({0.20})&-&-&-&{Gaia}\\
57983.44&{-6.31}&17.07(0.02)&17.24(0.01)&17.33(0.02)&17.63(0.02)&GROND\\
57984.14&{-5.67}&-&17.23(0.06)&17.20(0.05)&-&{LCO+Sinistro}\\
57986.80&{-3.22}&17.31(0.01)&17.23(0.01)&17.27(0.01)&-&{LCO+Sinistro}\\
57991.14&{0.78}&17.27(0.01)&17.27(0.02)&17.26(0.01)&-&{LCO+Sinistro}\\
{57991.15}&{0.78}&-&-&-&{17.80(0.05)}&{{LCO+Sinistro}}\\
57994.78&{4.13}&17.37(0.01)&17.21(0.01)&17.32(0.01)&-&{LCO+Sinistro}\\
57999.39&{8.36}&-&-&-&17.48(0.03)&{LCO+Sinistro}\\
{58003.38}&{12.03}&-&-&-&{17.48(0.03)}&{LCO+Sinistro}\\
58007.78&{16.08}&17.43(0.06)&17.44(0.03)&17.39(0.07)&-&{LCO+Sinistro}\\
58008.38&{16.63}&17.43(0.01)&17.39(0.01)&17.39(0.01)&17.41(0.01)&GROND\\
58017.37&{24.90}&17.99(0.01)&17.65(0.01)&17.77(0.01)&17.75(0.01)&GROND\\
58019.68&{27.03}&18.27(0.04)&17.91(0.04)&-&-&{LCO+Sinistro}\\
58020.36&{27.65}&18.18(0.01)&17.66(0.01)&17.78(0.01)&17.74(0.03)&GROND\\
58024.08&{31.08}&18.20(0.05)&17.98(0.02)&17.82(0.03)&-&{LCO+Sinistro}\\
58025.30&{32.20}&18.19(0.13)&18.06(0.04)&18.01(0.03)&17.77(0.06)&GROND\\
58027.12&{33.88}&18.37(0.05)&18.04(0.02)&17.86(0.03)&-&{LCO+Sinistro}\\
58029.32&{35.89}&18.39(0.02)&17.97(0.01)&17.75(0.03)&17.80(0.04)&GROND\\
58031.00&{37.45}&18.53(0.08)&18.09(0.06)&17.97(0.04)&-&{LCO+Sinistro}\\
58032.08&{38.43}&18.70(0.02)&18.27(0.03)&-&-&{LCO+Sinistro}\\
58033.30&{39.56}&18.88(0.02)&18.07(0.01)&18.04(0.02)&17.95(0.01)&GROND\\
58036.28&{42.30}&18.86(0.03)&-&18.13(0.02)&-&{LCO+Sinistro}\\
58040.02&{45.74}&18.89(0.02)&18.39(0.02)&18.28(0.02)&-&{LCO+Sinistro}\\
58040.24&{45.94}&18.94(0.01)&18.41(0.01)&18.37(0.01)&18.18(0.02)&GROND\\
58044.01&{49.41}&19.36(0.04)&18.72(0.03)&18.69(0.04)&-&{LCO+Sinistro}\\
58044.34&{49.71}&19.36(0.01)&18.53(0.04)&18.69(0.02)&18.59(0.04)&GROND\\
58047.75&{52.85}&19.62(0.06)&19.03(0.12)&-&-&{LCO+Sinistro}\\
58047.75&{52.85}&-&-&18.94(0.20)&-&{LCO+Sinistro}\\
58049.35&{54.32}&20.14(0.04)&19.32(0.04)&19.29(0.04)&18.77(0.05)&GROND\\
58054.29&{58.87}&20.50(0.04)&19.74(0.02)&19.54(0.03)&19.27(0.03)&GROND\\
58057.70&{62.01}&20.48(0.08)&19.90(0.06)&-&-&{LCO+Sinistro}\\
58063.21&{67.07}&20.60(0.11)&19.76(0.07)&19.52(0.04)&19.35(0.06)&GROND\\
58063.22&{67.09}&-&19.87(0.09)&19.59(0.08)&-&{LCO+Sinistro}\\
58068.31&{71.77}&20.59(0.05)&19.73(0.02)&19.50(0.02)&19.32(0.03)&GROND\\
58071.24&{74.46}&20.70(0.06)&20.01(0.05)&19.70(0.05)&-&{LCO+Sinistro}\\
58072.30&{75.44}&20.70(0.05)&19.77(0.02)&19.72(0.05)&19.33(0.04)&GROND\\
58078.20&{80.86}&20.76(0.07)&19.95(0.02)&19.77(0.02)&19.31(0.03)&GROND\\
58079.16&{81.75}&20.78(0.06)&20.13(0.05)&19.82(0.07)&-&{LCO+Sinistro}\\
\hline
\end{tabular}
\end{table*}
\begin{table*}
\caption{(continued)}
\begin{tabular}{ccccccc}
\hline
58083.23&{85.49}&20.81(0.04)&20.04(0.02)&19.75(0.02)&19.07(0.02)&GROND\\
58087.26&{89.20}&20.89(0.06)&20.00(0.02)&19.88(0.04)&19.46(0.03)&GROND\\
58092.19&{93.74}&20.95(0.10)&20.18(0.06)&20.01(0.06)&19.77(0.07)&GROND\\
58095.02&{96.34}&20.92(0.10)&20.28(0.11)&20.12(0.09)&-&{LCO+Sinistro}\\
58098.12&{99.19}&21.20(0.06)&20.34(0.02)&20.14(0.03)&19.92(0.05)&GROND\\
58101.16&{101.99}&21.17(0.06)&20.50(0.05)&20.05(0.06)&-&{LCO+Sinistro}\\
58102.20&{102.95}&21.07(0.06)&20.48(0.03)&19.89(0.02)&19.76(0.04)&GROND\\
58109.64&{109.79}&21.00(0.08)&20.34(0.06)&20.05(0.07)&-&{LCO+Sinistro}\\
58117.16&{116.70}&20.91(0.06)&20.16(0.05)&19.89(0.03)&-&{LCO+Sinistro}\\
58123.25&{122.31}&21.41(0.10)&20.60(0.03)&20.55(0.03)&20.04(0.06)&GROND\\
58124.90&{123.83}&21.43(0.11)&20.89(0.11)&20.95(0.13)&-&{LCO+Sinistro}\\
58129.25&{127.83}&21.86(0.12)&21.00(0.05)&20.86(0.05)&20.44(0.04)&GROND\\
58134.15&{132.34}&22.13(0.12)&21.20(0.05)&21.10(0.04)&20.52(0.05)&GROND\\
58139.20&{136.98}&22.28(0.12)&21.13(0.05)&21.23(0.05)&20.53(0.06)&GROND\\
58143.18&{140.64}&22.23(0.12)&21.24(0.06)&21.10(0.04)&20.74(0.07)&GROND\\
58151.10&{147.93}&21.72(0.12)&20.93(0.04)&20.74(0.03)&20.37(0.04)&GROND\\
58163.17&{159.03}&22.03(0.06)&21.21(0.05)&21.01(0.04)&20.70(0.04)&GROND\\
58170.12&{165.43}&22.22(0.19)&21.30(0.04)&21.16(0.05)&20.57(0.08)&GROND\\
58183.07&{177.34}&22.63(0.10)&21.72(0.06)&-&-&GROND\\
58188.09&{181.96}&-&22.14(0.06)&22.62(0.07)&21.76(0.07)&GROND\\
58191.18&{184.80}&-&22.34(0.05)&22.39(0.09)&21.67(0.09)&GROND\\
58192.04&{185.59}&23.25(0.10)&22.43(0.05)&23.17(0.09)&22.27(0.13)&GROND\\
58200.01&{192.93}&-&22.56(0.10)&23.17(0.12)&-&GROND\\
58206.01&{198.45}&-&22.56(0.09)&-&22.73(0.18)&GROND\\
58212.01&{203.97}&-&22.81(0.12)&22.81(0.18)&22.13(0.18)&GROND\\
58221.02&{212.25}&-&22.76(0.07)&-&-&GROND\\
58257.99&{246.27}&-&{$\gtrsim22.86$}&{$\gtrsim22.90$}&-&EFOSC2\\
58370.37&{349.65}&-&{$\gtrsim23.19$}&-&-&EFOSC2\\
58467.31&{438.83}&-&-&-&-&EFOSC2\\
58469.26&{440.63}&-&-&{$\gtrsim23.25$}&-&EFOSC2\\
\hline
\end{tabular}
\end{table*}
\begin{table*}
\centering
\caption{$U,B,V$-observed (non K-corrected, {non S-corrected}) magnitudes (in AB system). {\textit{Swift}/UVOT photometry was measured with a 5\arcsec-radius aperture (see text)}. Errors are in parentheses.}
\label{tab:bvtab}
\begin{tabular}{cccccc}
\hline
MJD&r. f. phase from maximum&$U$&$B$&$V$&instrument\\
\hline
57984.12&{-5.68}&-&17.36(0.05)&17.30(0.05)&{LCO+Sinistro}\\
57986.79&{-3.23}&-&17.23(0.01)&17.45(0.01)&{LCO+Sinistro}\\
57987.78&{-2.31}&-&17.33(0.01)&17.42(0.00)&{LCO+Sinistro}\\
57989.82&{-0.44}&17.48(0.05)&-&-&{\textit{Swift}/UVOT}\\
57989.82&{-0.44}&-&17.21(0.06)&-&{\textit{Swift}/UVOT}\\
57989.82&{-0.44}&-&-&17.21(0.12)&{\textit{Swift}/UVOT}\\
57991.14&{0.77}&-&17.07(0.03)&17.43(0.01)&{LCO+Sinistro}\\
57994.78&{4.12}&-&17.36(0.01)&17.33(0.00)&{LCO+Sinistro}\\
57998.29&{7.35}&17.76(0.09)&-&-&{\textit{Swift}/UVOT}\\
57998.29&{7.35}&-&17.34(0.10)&-&{\textit{Swift}/UVOT}\\
57998.29&{7.35}&-&-&17.17(0.16)&{\textit{Swift}/UVOT}\\
57999.34&{8.32}&-&17.43(0.02)&17.38(0.01)&{LCO+Sinistro}\\
58001.47&{10.28}&17.92(0.06)&-&-&{\textit{Swift}/UVOT}\\
58001.47&{10.28}&-&17.39(0.06)&-&{\textit{Swift}/UVOT}\\
58001.48&{10.28}&-&-&17.28(0.11)&{\textit{Swift}/UVOT}\\
58003.12&{11.79}&-&17.51(0.01)&17.25(0.01)&{LCO+Sinistro}\\
58007.10&{15.46}&-&17.74(0.00)&17.32(0.01)&{LCO+Sinistro}\\
58007.76&{16.06}&-&17.80(0.01)&17.37(0.01)&{LCO+Sinistro}\\
58011.73&{19.71}&-&17.85(0.01)&17.56(0.01)&{LCO+Sinistro}\\
58015.68&{23.35}&-&18.27(0.01)&17.68(0.01)&{LCO+Sinistro}\\
58027.12&{33.87}&-&18.74(0.01)&18.18(0.01)&{LCO+Sinistro}\\
58031.00&{37.44}&-&18.76(0.04)&18.19(0.04)&{LCO+Sinistro}\\
58032.06&{38.42}&-&18.70(0.01)&18.19(0.01)&{LCO+Sinistro}\\
58036.26&{42.29}&-&18.91(0.02)&18.39(0.01)&{LCO+Sinistro}\\
58040.00&{45.72}&-&19.31(0.01)&18.56(0.01)&{LCO+Sinistro}\\
58044.00&{49.40}&-&19.68(0.04)&18.82(0.05)&{LCO+Sinistro}\\
58047.74&{52.84}&-&20.25(0.02)&19.50(0.02)&{LCO+Sinistro}\\
58057.70&{62.00}&-&20.92(0.08)&20.20(0.06)&{LCO+Sinistro}\\
58071.22&{74.44}&-&21.14(0.06)&20.02(0.03)&{LCO+Sinistro}\\
58095.01&{96.33}&-&21.56(0.10)&20.26(0.03)&{LCO+Sinistro}\\
58101.14&{101.97}&-&21.72(0.06)&20.68(0.04)&{LCO+Sinistro}\\
58109.62&{109.77}&-&21.16(0.08)&20.59(0.06)&{LCO+Sinistro}\\
58117.14&{116.68}&-&21.20(0.06)&20.41(0.03)&{LCO+Sinistro}\\
58124.88&{123.80}&-&21.82(0.09)&20.86(0.05)&{LCO+Sinistro}\\
\hline
\end{tabular}
\end{table*}
\begin{table*}
\centering
\caption{{NIR-observed (non K-corrected) template-subtracted ($J,H$) magnitudes and PSF ($K_{\rm s}$) magnitudes} (in AB system). Errors are in parentheses.}
\label{tab:jhktab}
\begin{tabular}{cccccc}
\hline
MJD&r. f. phase from maximum&$J$&$H$&$K_\mathrm{s}$&instrument\\
\hline
57983.44&{-6.31}&{17.75}({0.02})&{18.19}({0.03})&18.68(0.08)&GROND\\
57996.40&{5.61}&{17.72}({0.05})&{$\gtrsim14.957$}&18.64(0.12)&SOFI\\
58008.38&{16.63}&{17.82}({0.02})&{18.13}({0.04})&18.39(0.05)&GROND\\
58017.37&{24.90}&{17.99}({0.04})&{18.40}({0.05})&18.73(0.09)&GROND\\
58017.38&{24.91}&{17.92}({0.02})&{18.23}({0.03})&18.63(0.05)&SOFI\\
58020.36&{27.65}&{18.16}({0.02})&{18.30}({0.04})&18.66(0.05)&GROND\\
58025.30&{32.20}&{17.95}({0.07})&{18.42}({0.09})&-&GROND\\
58029.31&{35.89}&{17.99}({0.03})&{18.50}({0.03})&18.80(0.07)&GROND\\
58033.30&{39.56}&{18.16}({0.02})&{18.57}({0.04})&18.94(0.08)&GROND\\
58056.29&{60.71}&{$\gtrsim17.64$}&{19.46}({0.04})&19.70(0.06)&SOFI\\
58078.20&{80.86}&{19.70}({0.03})&{19.52}({0.04})&-&GROND\\
58083.23&{85.49}&{19.29}({0.04})&{19.51}({0.05})&$\gtrsim19.76$&GROND\\
58087.26&{89.20}&{19.44}({0.04})&{19.45}({0.04})&19.38(0.10)&GROND\\
58092.19&{93.74}&{19.47}({0.06})&{19.48}({0.09})&${\gtrsim19.52}$&GROND\\
58098.12&{99.19}&{19.80}({0.08})&{19.63}({0.05})&{19.38(0.10)}&GROND\\
58102.21&{102.95}&-&{19.51}({0.06})&-&GROND\\
58108.11&{108.38}&{19.64}({0.04})&{19.65}({0.06})&{$\gtrsim19.50$}&GROND\\
58118.23&{117.69}&{19.59}({0.07})&{19.27}({0.04})&{19.93(0.19)}&GROND\\
58123.25&{122.31}&{20.04}({0.09})&{19.74}({0.08})&-&GROND\\
58129.25&{127.83}&{20.67}({0.09})&{20.26}({0.09})&{$\gtrsim18.98$}&GROND\\
58134.15&{132.34}&{20.55}({0.08})&{19.98}({0.10})&19.37(0.17)&GROND\\
58139.20&{136.98}&{20.54}({0.13})&{20.32}({0.08})&${\gtrsim19.91}$&GROND\\
58143.18&{140.64}&{20.52}({0.06})&{20.66}({0.08})&-&GROND\\
58163.17&{159.03}&{20.44}({0.09})&{19.86}({0.07})&{$\gtrsim19.29$}&GROND\\
58170.12&{165.43}&{20.65}({0.12})&-&-&GROND\\
58178.23&{172.89}&{$\gtrsim20.40$}&{20.04}({0.11})&{$\gtrsim19.09$}&GROND\\
58183.07&{177.34}&{$\gtrsim20.63$}&{20.35}({0.16})&{$\gtrsim19.06$}&GROND\\
58188.09&{181.96}&{$\gtrsim21.03$}&{$\gtrsim20.62$}&-&GROND\\
58191.18&{184.80}&{$\gtrsim21.05$}&{$\gtrsim20.50$}&{$\gtrsim19.44$}&GROND\\
58192.04&{185.59}&{$\gtrsim19.49$}&{$\gtrsim20.56$}&{$\gtrsim18.14$}&GROND\\
58200.01&{192.93}&{$\gtrsim21.69$}&-&{$\gtrsim19.59$}&GROND\\
58206.01&{198.45}&{$\gtrsim21.58$}&{$\gtrsim20.91$}&{$\gtrsim19.43$}&GROND\\
58212.01&{203.97}&{21.43}({0.15})&{$\gtrsim21.05$}&{$\gtrsim19.23$}&GROND\\
58221.02&{212.25}&-&{20.74}({0.09})&-&GROND\\
58411.32&{387.32}&{21.78}({0.13})&{$\gtrsim21.25$}&{$\gtrsim20.22$}&GROND\\
58427.25&{401.98}&{$\gtrsim21.73$}&{$\gtrsim21.41$}&{$\gtrsim20.22$}&GROND\\
\hline
\end{tabular}
\end{table*}
\begin{table*}
\centering
\caption{S-corrections for GROND bands.}
\label{tab:scorrgrond}
\begin{tabular}{cccc}
\hline
{MJD}&{$g$}&{$r$}&{$i$}\\
\hline
{57892.39}&{-0.001}&{-0.013}&{-0.007}\\
{57984.39}&{0.053}&{-0.002}&{-0.030}\\
{57986.38}&{0.045}&{-0.007}&{-0.043}\\
{57987.38}&{0.047}&{-0.009}&{-0.040}\\
{58025.31}&{-0.017}&{0.028}&{0.020}\\
{58045.28}&{-0.057}&{0.008}&{0.022}\\
{58069.21}&{-0.081}&{0.003}&{0.032}\\
{58102.20}&{-0.066}&{-0.010}&{0.037}\\
{58132.45}&{-0.095}&{-0.031}&{0.054}\\
{58135.22}&{-0.075}&{-0.030}&{0.036}\\
{58159.23}&{-0.102}&{0.0560}&{-0.074}\\
{58192.10}&{-0.045}&{0.0570}&{-0.052}\\
\hline
\end{tabular}
\end{table*}
\begin{table*}
\centering
\caption{S-corrections for Sinistro filters.}
\label{tab:scorrlco}
\begin{tabular}{ccccccc}
\hline
{MJD}&{$B$}&{$g$}&{$V$}&{$r$}&{$i$}&{$z$}\\
\hline
{57892.39}&{0.000}&{0.001}&{0.004}&{-0.016}&{-0.009}&{-0.005}\\
{57984.39}&{-0.002}&{-0.008}&{0.036}&{-0.006}&{-0.009}&{-0.059}\\
{57986.38}&{-0.002}&{-0.006}&{0.024}&{-0.005}&{-0.013}&{-0.057}\\
{57987.38}&{-0.003}&{-0.007}&{0.026}&{-0.007}&{-0.012}&{-0.057}\\
{58025.31}&{-0.003}&{0.013}&{0.010}&{-0.007}&{0.000}&{-0.054}\\
{58045.28}&{-0.008}&{0.021}&{-0.004}&{-0.017}&{-0.003}&{-0.052}\\
{58069.21}&{-0.006}&{0.018}&{-0.013}&{-0.014}&{-0.003}&{-0.049}\\
{58102.20}&{-0.009}&{0.015}&{-0.012}&{-0.026}&{-0.005}&{-0.051}\\
{58132.45}&{-0.006}&{-0.001}&{-0.014}&{-0.027}&{-0.007}&{-0.009}\\
{58135.22}&{-0.009}&{0.010}&{-0.014}&{-0.037}&{-0.007}&{-0.049}\\
{58159.23}&{-0.013}&{0.024}&{-0.014}&{0.056}&{-0.007}&{-0.008}\\
{58192.10}&{0.001}&{0.053}&{-0.008}&{0.020}&{-0.010}&{-0.005}\\
\hline
\end{tabular}
\end{table*}
\begin{table*}
\centering
\caption{S-corrections for \textit{Swift}/UVOT filters.}
\label{tab:scorrswift}
\begin{tabular}{cccc}
\hline
{MJD}&{$B$}&{$V$}\\
\hline
{57892.39}&{-0.000}&{0.013}\\
{57984.39}&{0.001}&{0.019}\\
{57986.38}&{-0.001}&{0.010}\\
{57987.38}&{-0.001}&{0.012}\\
{58025.31}&{0.006}&{0.049}\\
{58045.28}&{0.001}&{0.037}\\
{58069.21}&{-0.001}&{0.019}\\
{58102.20}&{-0.001}&{0.003}\\
{58132.45}&{-0.012}&{-0.017}\\
{58135.22}&{-0.011}&{-0.010}\\
{58159.23}&{-0.024}&{-0.004}\\
{58192.10}&{-0.000}&{0.020}\\
\hline
\end{tabular}
\end{table*}
\begin{table*}
    \centering
        \caption{Estimated uncertainties $\Delta S_{\rm corr}$ for the filters $U,z,J,H,K_{\rm s}$ (for each instrument) divided in two temperature ranges (see text). }
        \label{tab:scorrerr}
    \begin{tabular}{ccc}
    
    \hline
         &{$4000\,\mathrm{K}<T<8000\,\mathrm{K}$}&{$8000\,\mathrm{K}<T<12000\,\mathrm{K}$}  \\
         \hline
         {GROND}&\begin{tabular}{l}{$\Delta S_{{\rm corr,}z}=0.010\,\mathrm{mag}$}\\{$\Delta S_{{\rm corr,}J}=0.004\,\mathrm{mag}$}\\{$\Delta S_{{\rm corr,}H}=0.001\,\mathrm{mag}$}\\{$\Delta S_{{\rm corr,}K_{\rm s}}=0.010\,\mathrm{mag}$}\end{tabular}&\begin{tabular}{l}{$\Delta S_{{\rm corr,}z}=0.002\,\mathrm{mag}$}\\{$\Delta S_{{\rm corr,}J}=0.001\,\mathrm{mag}$}\\{$\Delta S_{{\rm corr,}H}=0.000\,\mathrm{mag}$}\\{$\Delta S_{{\rm corr,}K_{\rm s}}=0.001\,\mathrm{mag}$}\end{tabular}\\
         \hline
         {SOFI}&\begin{tabular}{l}{$\Delta S_{{\rm corr,}J}=0.020\,\mathrm{mag}$}\\{$\Delta S_{{\rm corr,}H}=0.080\,\mathrm{mag}$}\\{$\Delta S_{{\rm corr,}K_{\rm s}}=0.070\,\mathrm{mag}$}\end{tabular}&\begin{tabular}{l}{$\Delta S_{{\rm corr,}J}=0.002\,\mathrm{mag}$}\\{$\Delta S_{{\rm corr,}H}=0.001\,\mathrm{mag}$}\\{$\Delta S_{{\rm corr,}K_{\rm s}}=0.120\,\mathrm{mag}$}\end{tabular}\\
         \hline
         {LCO+Sinistro}&\begin{tabular}{l}{$\Delta S_{{\rm corr,}z}=0.001\,\mathrm{mag}$}\end{tabular}&\begin{tabular}{l}{$\Delta S_{{\rm corr,}z}=0.005\,\mathrm{mag}$}\end{tabular}\\
         \hline
         {\textit{Swift}/UVOT}&\begin{tabular}{l}{$\Delta S_{{\rm corr,}U}=0.2\,\mathrm{mag}$}\end{tabular}&\begin{tabular}{l}{$\Delta S_{{\rm corr,}U}=0.05\,\mathrm{mag}$}\end{tabular}\\
         \hline
    \end{tabular}
    \label{tab:my_label}
\end{table*}
\begin{table*}
\centering
\caption{K-corrections expressed in magnitudes.}
\label{tab:kcorr}
\begin{tabular}{cccccccccccccc}
\hline
r. f. phase from maximum&$UVW2$&$UVM2$&$UVW1$&$U$&$B$&$g$&$V$&$r$&$i$&$z$&$J$&$H$&$K_\mathrm{s}$\\
\hline
{-7}&0.39&0.51&0.26&0.04&-0.18&-0.18&-0.21&-0.23&-0.24&-1.55&-0.24&-0.25&-0.25\\
{-5}&0.37&0.5&0.26&0.04&-0.17&-0.17&-0.21&-0.24&-0.24&-0.24&-0.22&-0.22&-0.25\\
{-4}&0.35&0.49&0.26&0.04&-0.16&-0.15&-0.17&-0.16&-0.27&-1.45&-0.22&-0.23&-0.24\\
{32}&0.06&0.28&0.27&0.02&0.08&0.02&-0.07&-0.21&-0.14&-1.54&-0.16&-0.20&-0.22\\
{51}&-0.07&0.17&0.27&0.02&0.22&0.13&-0.02&-0.13&-0.16&-1.52&-0.15&-0.18&-0.21\\
{73}&-0.24&0.02&0.28&0.01&0.30&0.19&0.01&-0.08&-0.06&-1.60&0.58&-0.10&-0.20\\
{103}&-0.4&-0.13&0.28&0.01&0.26&0.15&-0.01&-0.05&-0.09&-1.45&0.18&-0.16&-0.19\\
135&0.244&0.179&0.006&0.023&-0.094&-1.455&        &       &         \\
+175&0.164&0.097&-0.067&0.097&-0.131&0.096&1.64097&-0.16296&-0.19509\\
+358&0.106&0.066&-0.085&0.066&-0.17&-0.378&-0.11968&-0.1644&-0.19602\\
\hline
\end{tabular}
\end{table*}
\begin{table*}
\centering
\caption{Slopes of the observed LCs [$10^{-2}$ mag/day]}
\label{tab:slopes}
\begin{tabular}{ccccccccccccc}
\hline
$UVW2$&$UVM2$&$UVW1$&$U$&$B$&$g$&$V$&$r$&$i$&$z$&$J$&$H$&$K_\mathrm{s}$\\
\hline
-&-&-&-&-&{2.22}&-&2.23&2.28&2.25&{2.23}&{1.80}&-\\
\hline
\end{tabular}
\end{table*}

\begin{table*}
\centering
\caption{Spectra in Fig.~\ref{fig:spec_evol}.}
\label{tab:sfo}
\begin{tabular}{ccccc}
\hline
MJD&r. f. phase from maximum&instrument&resolution\\
\hline
57982.39 & {-7} & EFOSC2 & 18.2 \\
57984.39 & {-5} & EFOSC2 & 17.9 \\
57986.38 & {-4} & EFOSC2 & 18.9 \\
57987.38 & {-3} & EFOSC2 & 27.2 \\
{58011.73} & {20} & {LCO+FLOYDS} & {21.1} \\
{58015.73} & {23} & {LCO+FLOYDS} & {21.8} \\
{58021.73} & {29} & {LCO+FLOYDS} & {22.1} \\
58025.31 & {32} & EFOSC2 & 17.9 \\
{58030.71} & {37} & {LCO+FLOYDS} & {20.0} \\
{58035.73} & {42} & {LCO+FLOYDS} & {18.4} \\
58045.28 & {51} & EFOSC2 & 17.8 \\
58069.21 & {73} & EFOSC2 & 17.8 \\
58102.20 & {103} & EFOSC2 & 17.8 \\
58132.45 & {131} & LRIS & - \\
58135.22 & {133} & EFOSC2 & 18.0 \\
58159.23 & {155} & Binospec & - \\
58192.10 & {187} & X-Shooter & - \\
58389.35 & {367} & X-Shooter & - \\
\hline
\end{tabular}
\end{table*}

\begin{table*}
\centering
\caption{Logarithm of the bolometric luminosities integrated over the $UVW2,UVM2,UVW1,U,B,g,V,r,i,z,J,H,K_\mathrm{s}$, the blackbody temperatures (expressed in Kelvin). Errors are reported in parenthesis. {We fixed a maximum error for the blackbody temperatures to 2000 K (see text). Epochs later than $\sim$160 days require even larger errorbars.}}
\label{tab:blc}
\begin{tabular}{ccc}
\hline
r. f. phase from maximum&$\log_{10}L_\mathrm{bol}$& $T_\mathrm{BB}$\\
\hline
{-5.57}&{43.65(0.06)}&{11693.01}\\
{-4.94}&{43.63(0.06)}&{11487.76}\\
{-2.48}&{43.67(0.07)}&{11633.86}\\
{1.51}&{43.75(0.07)}&{11429.11}\\
{4.86}&{43.77(0.07)}&{10699.96}\\
{16.81}&{43.7(0.06)}&{10775.08}\\
{17.37}&{43.73(0.06)}&{9313.39}\\
{25.64}&{43.61(0.04)}&{8909.54}\\
{27.76}&{43.53(0.04)}&{8597.25}\\
{28.39}&{43.58(0.04)}&{8393.45}\\
{31.81}&{43.5(0.04)}&{8351.5}\\
{32.93}&{43.46(0.04)}&{8555.91}\\
{34.61}&{43.47(0.04)}&{8681.41}\\
{36.62}&{43.48(0.05)}&{8248.12}\\
{38.18}&{43.44(0.05)}&{8210.32}\\
{39.16}&{43.44(0.04)}&{7489.47}\\
{40.29}&{43.45(0.04)}&{7478.19}\\
{46.47}&{43.33(0.03)}&{6997.47}\\
{46.68}&{43.33(0.03)}&{6997.47}\\
{50.15}&{43.21(0.03)}&{6887.7}\\
{50.45}&{43.2(0.03)}&{6205.62}\\
{53.59}&{43.03(0.04)}&{6061.82}\\
{55.06}&{42.98(0.04)}&{5678.73}\\
{59.6}&{42.87(0.03)}&{5640.1}\\
{62.74}&{42.85(0.05)}&{5586.08}\\
{67.81}&{42.84(0.04)}&{5562.2}\\
{67.82}&{42.84(0.04)}&{5421.8}\\
{72.5}&{42.84(0.03)}&{5471.99}\\
{75.19}&{42.8(0.03)}&{5356.28}\\
{76.17}&{42.82(0.03)}&{5357.45}\\
{81.6}&{42.78(0.03)}&{5407.49}\\
{82.48}&{42.78(0.03)}&{5324.42}\\
{86.23}&{42.79(0.03)}&{5218.12}\\
{89.94}&{42.77(0.03)}&{5192.4}\\
{94.47}&{42.74(0.03)}&{4961.57}\\
{97.07}&{42.73(0.04)}&{4920.14}\\
{99.93}&{42.66(0.03)}&{4974.52}\\
{102.72}&{42.63(0.03)}&{5085.91}\\
{103.68}&{42.65(0.02)}&{5171.26}\\
{110.52}&{42.68(0.03)}&{4629.46}\\
{117.43}&{42.72(0.03)}&{4610.89}\\
{123.05}&{42.54(0.04)}&{4338.4}\\
{124.56}&{42.52(0.05)}&{4165.87}\\
{128.56}&{42.43(0.06)}&{4180.01}\\
{133.07}&{42.39(0.05)}&{4121.38}\\
{137.72}&{42.39(0.06)}&{4384.67}\\
{141.38}&{42.27(0.12)}&{4236.11}\\
{148.67}&{42.44(0.04)}&{4162.07}\\
{159.77}&{42.39(0.04)}&{3710.72}\\
{166.16}&{42.53(0.07)}&-\\
{178.08}&{42.31(0.08)}&-\\  
{185.54}&{42.28(0.07)}&-\\  
{186.33}&{42.13(0.09)}&-\\  
{193.66}&{42.17(0.13)}&-\\  
{199.18}&{42.11(0.14)}&-\\  
{204.7}&{42.16(0.1)}&-\\
{212.99}&{42.16(0.1)}&-\\
\hline
\end{tabular}
\end{table*}
\begin{table*}
\centering
\caption{Comparison of the metallicity estimated for SN 2017gci with the metallicities of a sample of nearby SLSNe~I and GRBs \citep[data from][]{chenetal2017b}.}
\label{tab:metall}
\begin{tabular}{llllllll}
\hline
Object&SN 2017gci&SN 2017egm&PTF11hrq&PTF12dam&GRB 100316D&GRB 060505&GRB 111005A\\
\hline
Reference&Sec. \ref{sec:metall}&\citep{chenetal2017b}&\citep{cikotaetal2017}&\citep{thoeneetal2015}&\citep{izzoetal2018}&\citep{thoeneetal2014}&\citep{tangaetal2017}\\
Redshift&0.087&0.031&0.057&0.107&0.059&0.089&0.013\\
PP04 O3N2&$8.135\pm0.07$&$8.77\pm0.01$&$8.19\pm0.01$&$8.01\pm0.14$&$8.21\pm0.02$&$8.24\pm0.00$&$8.63\pm0.03$\\
\hline
\end{tabular}
\end{table*}
\begin{table*}
\centering
\caption{Best-fit estimates of the physical parameters of SN~2017gci (with reference to Fig.~\ref{fig:blc}). {The \texttt{TigerFit} best-fit model is listed in the first column with the assumed phase from the explosions in square brackets.}}
\label{tab:fp}
\begin{tabular}{ccccccccccc}
\hline
&ejecta&{mass}&polar mag.&initial&phase from the&opacity&{CSM}&{progenitor}&diffusion&spin-down\\
&mass $M_\mathrm{ejecta}$&{accretion rate}&field $B_\mathrm{p}$&period $P_\mathrm{initial}$&{explosion} $\phi_0$&$\kappa$&{mass}&{radius}&timescale&timescale\\
&$[\mathrm{M}_\odot]$&{$[\mathrm{M}_\odot\,\mathrm{year}^{-1}]$}&$[10^{14}\,\mathrm{G}]$&$[\mathrm{ms}]$&$[\mathrm{days}]$&$[\mathrm{cm^2\,g^{-1}}]$&{[$\mathrm{M}_\odot$]}&{[$10^{14}$ cm]}&[days]&[days]\\
\hline
MF1&{9.0}&-&{5.5}&{2.8}&{15.7}&{0.08}&-&-&{34.5}&{1.1}\\
{\texttt{csm0} [30]}&{12.4}&{0.1}&-&-&-&{0.07}&{4.9}&{0.004}&-&-\\
\hline
\end{tabular}
\end{table*}

\bsp	
\label{lastpage}
\end{document}